\documentclass[twocolumn,trackchanges]{aastex701}

\usepackage{subcaption}    % 提供 subfigure 环境
\usepackage{amsmath,amssymb,mathtools} % 如果不放心，可以保留
\usepackage{bm}
\usepackage{float}
\usepackage{romannum}
\newcommand{\rom}[1]{%
  \ifcase#1\or I\or II\or III\or IV\or V\or VI\or VII\or VIII\or IX\or X\fi}

\begin{document}

% \title{Machine Learning-Based Estimation of Lensing Amplification Factors in the Gravitational-Wave Frequency Band}
\title{Efficient Evaluation of Gravitational Lensing Amplification Factors: A Deep Learning Framework}

% \title{\textbf{Machine Learning-Based Estimation of Amplification Factor in the Presence of Point-Mass Lensing}}% 

\author{Fan Zhang}
\altaffiliation{These authors contributed equally to this work.}
\email{fzhang@zju.edu.cn}  
\affiliation{%
 State Key Laboratory of Ocean Sensing {\normalfont\&} Ocean College, Zhejiang University, Zhoushan, 316021, China
}%
\affiliation{%
 Kavli Institute for Astrophysics and Space Research, Massachusetts Institute of Technology, Cambridge 02139, MA, USA
}%

\author{Qikai Zhang}%
\altaffiliation{These authors contributed equally to this work.}
\email{15895935591@163.edu.cn}  
\affiliation{%
 State Key Laboratory of Ocean Sensing {\normalfont\&} Ocean College, Zhejiang University, Zhoushan, 316021, China
}%

% \author{Fan Zhang}
% \email{fzhang@zju.edu.cn}  
% \affiliation{%
%  State Key Laboratory of Ocean Sensing {\normalfont\&} Ocean College, Zhejiang University, Zhoushan, 316021, China
% }%
% \affiliation{%
%  Kavli Institute for Astrophysics and Space Research, Massachusetts Institute of Technology, Cambridge 02139, MA, USA
% }

% \author{Qikai Zhang}%
% \email{15895935591@163.edu.cn}  
% \affiliation{%
%  State Key Laboratory of Ocean Sensing {\normalfont\&} Ocean College, Zhejiang University, Zhoushan, 316021, China
% }%

\author{Qiyuan Yang}
\email{yangqiyuan@whu.edu.cn}  
 % \homepage{http://www.Second.institution.edu/~Charlie.Author}
\affiliation{
 School of Physics and Technology, Wuhan University, Wuhan 430072, China
}%

\author[orcid=0009-0002-9540-9230]{Yong Yuan}
\email{yuanyong@imech.ac.cn}
\affiliation{Center for Gravitational Wave Experiment, National Microgravity Laboratory, Institute of Mechanics, Chinese Academy of Sciences, Beijing, China}

\author[orcid=0000-0002-8174-0128]{Xilong Fan}
 % \homepage{http://www.Second.institution.edu/~Charlie.Author}
\email[show]{Contact author: xilong.fan@whu.edu.cn}
\affiliation{
 School of Physics and Technology, Wuhan University, Wuhan 430072, China
}%
\correspondingauthor{Xilong Fan}

% \author{Delta Author}
% \affiliation{%
%  Authors' institution and/or address\\
%  This line break forced with \textbackslash\textbackslash
% }%

% \author{Fan Zhang}
% \affiliation{State Key Laboratory of Ocean Sensing \& Ocean College, Zhejiang University, Zhoushan, 316021, China}
% \affiliation{Kavli Institute for Astrophysics and Space Research, Massachusetts Institute of Technology, Cambridge, MA 02139, USA}
% \email{fzhang@zju.edu.cn}

% \author{Qikai Zhang}
% \affiliation{State Key Laboratory of Ocean Sensing \& Ocean College, Zhejiang University, Zhoushan, 316021, China}
% \email{15895935591@163.edu.cn}

% \author{Xilong Fan}
% \affiliation{School of Physics and Technology, Wuhan University, Wuhan 430072, China}
% \email{xilong.fan@whu.edu.cn}
% \thanks{Corresponding author: Xilong Fan}

% \collaboration{CLEO Collaboration}%\noaffiliation

\date{\today}% It is always \today, today,
             %  but any date may be explicitly specified

\begin{abstract}
Wave optics is essential for analyzing lensed gravitational waves (GWs), yet evaluating the diffraction integral $F(\omega, y)$ is computationally expensive.
We present a Sinusoidal Representation Networks (SIRENs) framework for the dimensionless amplification factor, demonstrating its efficacy and generalization through Point Mass Lens (PML) and Singular Isothermal Sphere (SIS) test cases.
Unlike standard architectures that suffer from spectral bias, the network's periodic activation functions structurally align with the integral's oscillatory kernel, effectively resolving high-frequency spectral features.
The resulting estimator achieves $\mathcal{O}(10^{-3})$ relative accuracy and a $\sim 100\times$ speedup compared to direct numerical integration.
By shifting the computational burden to offline training, our framework yields a stable $\mathcal{O}(1)$ inference complexity. This guarantees constant, sub-millisecond evaluation times even in the weak-lensing diffraction tail where traditional methods stagnate.
Additionally, the dimensionless formulation ensures intrinsic scale invariance, enabling direct application across astrophysical regimes from stellar-mass lenses in the ground-based LVK band to supermassive black holes in the space-based LISA band.
\end{abstract}

% \begin{abstract}
% Wave optics is essential for analyzing lensed gravitational waves (GWs), yet evaluating the diffraction integral $F(\omega, y)$ is computationally expensive.
% We present a Sinusoidal Representation Networks (SIRENs) framework for the dimensionless amplification factor, demonstrating its efficacy and generalization through Point Mass Lens (PML) and Singular Isothermal Sphere (SIS) test cases.
% The network's periodic activation functions structurally align with the integral's oscillatory kernel, effectively resolving high-frequency spectral features.
% The resulting estimator achieves $\mathcal{O}(10^{-3})$ relative accuracy and a $\sim 100\times$ speedup compared to direct numerical integration.
% Furthermore, the framework shifts the heavy computational burden to offline training, yielding a stable $\mathcal{O}(1)$ inference complexity that remains constant even in the highly oscillatory regime.
% Additionally, the dimensionless formulation ensures intrinsic scale invariance, enabling direct application across astrophysical regimes from stellar-mass lenses in the LVK band to supermassive black holes in the LISA band.
% \end{abstract}

%\keywords{Suggested keywords}%Use showkeys class option if keyword
                              %display desired
% \maketitle

\section{Introduction}

Gravitational lensing of GWs provides a new probe of compact objects and small-scale structure that are difficult to access electromagnetically. When the GW wavelength becomes comparable to the characteristic scale of the lens, diffraction and interference produce frequency-dependent modulations in the observed signal, known as the wave-optics regime \citep{Nakamura1998,takahashi2003wave,Deguchi1986}. Such signatures are expected from stellar-mass black holes, compact dark matter substructure and intermediate-mass black holes (IMBHs) lensing, and they are relevant to both current LVK~\citep{abbott2020prospects} ground-based detector network~\citep{aasi2015advanced,acernese2014advanced,kagra2019kagra} and future space missions LISA~\citep{amaro2017laser}, Taiji~\citep{luo2021taiji}, and TianQin~\citep{luo2016tianqin}. Within the ground-based GW detector band, we focus on lensed GWs from compact–binary coalescences (CBCs). In parts of this parameter space, the geometric–optics approximation remains adequate; for example, \(\sim\!100\,M_\odot\) PML can be well described by geometric optics for typical LIGO-band signals~\citep{liu2023exploring}. On the other hand, wave-optics effects become relevant near the diffraction–interference transition \citep{takahashi2003wave,Nakamura1998,refsdal1994gravitational,lai2018discovering,liao2018anomalies,christian2018detecting,pagano2020lensinggw,oguri2022amplitude,seo2022improving,wright2022gravelamps,basak2022constraints,tambalo2023lensing,liu2023exploring}. This effect becomes non-negligible for lens masses \(M_L \lesssim 10^3\,M_\odot\) within the LVK sensitivity band (\(\sim\!20\text{--}1000\) Hz) and likewise within the planned space-based detector LISA ~\citep{amaro2017laser} band (\(\sim\!10^{-4}\text{--}{1}\) Hz). Meanwhile, LVK searches in the O3 data have reported no confident lensed GW detections, highlighting the requirement for accurate and fast forward models of lensed GWs to enable the exhaustive parameter scans \citep{abbott2021search,abbott2023search,ligo2025gwtc}.  

Recent studies have clarified the phenomenology across strong lensing and microlensing, developed fast evaluators for wave-optics amplification, and highlighted astrophysical applications \citep{lai2018discovering,liao2018anomalies,christian2018detecting,pagano2020lensinggw,kim2021identification,bulashenko2022lensing,oguri2022amplitude,seo2022improving,wright2022gravelamps,basak2022constraints,tambalo2023lensing}. In the wave-optics regime, lensing is encoded by a complex amplification factor \(F(\omega,y)\) that maps the unlensed strain to the lensed one as a function of angular frequency \(\omega\) and the dimensionless source–lens offset \(y\) (in units of the Einstein radius). Even for the analytically tractable PML \citep{Deguchi1986}, \(F(\omega,y)\) exhibits highly oscillatory structure, particularly at moderate-to-large \(y\), which makes direct evaluation via special functions or diffraction integrals computationally demanding.
Considerable effort has been devoted to optimizing these numerical integrations. For instance, ~\cite{Guo2020} investigated the convergence of diffraction integrals and proposed a refined method combining zero-point integration with asymptotic expansions to improve both accuracy and efficiency.
Despite these improvements, direct numerical integration remains too slow for real-time generation of waveforms in large-scale parameter estimation pipelines, which may require \(\mathcal{O}(10^7)\) likelihood evaluations per event ~\citep{smith2020massively}.

Consequently, recent efforts have focused on bypassing these computational bottlenecks through optimized numerical codes, such as GLoW~\citep{villarrubia2024glow} and wolensing~\citep{yeung2024wolensing}. However, existing surrogate paradigms face distinct limitations. Likelihood-free tools like Conditional Variational Autoencoders ~\citep{nerin2025parameter} estimate posteriors directly but bypass the amplification factor entirely, precluding explicit waveform evaluation. Alternatively, time-domain emulators using Reduced Basis Methods~\citep{deka2025surrogate} introduce computational bottlenecks during frequency-domain transformation and require complex manual interventions to handle singularities. While direct frequency-domain deep learning offers a fully automated alternative, standard architectures like Multi-Layer Perceptrons (MLPs) suffer from spectral bias~\citep{rahaman2019spectral}, failing to resolve the dense high-frequency interference fringes characteristic of wave optics.

% Consequently, recent research has focused on developing faster approximate methods or surrogate models.
% On the numerical side, optimized codes such as GLoW~\citep{villarrubia2024glow} and wolensing~\citep{yeung2024wolensing} have been introduced to accelerate wave-optics computations.
% On the machine learning front, neural networks have emerged as promising surrogates.
% For example, ~\cite{nerin2025parameter} utilized conditional variational autoencoders for parameter estimation of microlensed GWs, and ~\cite{deka2025surrogate} applied reduced basis methods to build surrogates for spherically symmetric potentials.
% However, standard neural networks, such as Multi-Layer Perceptrons (MLPs) with ReLU activations, suffer from spectral bias, which refers to the tendency to learn low-frequency features while struggling to capture high-frequency oscillations~\citep{rahaman2019spectral}.

In this work, we employ SIRENs~\citep{sitzmann2020implicit} to directly model the amplification factor \(F(\omega,y)\) in the frequency domain.
While neural networks are inherently ``black box'' models, the specific architectural choice of SIRENs provides a essential advantage: their periodic activation functions possess an structural correspondence that is mathematically isomorphic to the oscillatory nature of the diffraction integral.
This allows us to circumvent the spectral bias and accurately resolve dense interference fringes without determining time-domain counterparts. To demonstrate the efficacy of our approach, we focus on the phenomenologically rich transition zone of the wave-optics regime.
This domain spans from the diffraction-dominated limit to the regime of significant interference~\citep{takahashi2003wave, Nakamura1998}, representing a critical benchmark where the complex, dense oscillatory fringes render traditional geometric optics approximations inaccurate~\citep{liu2023exploring}.
This regime requires full wave-optics calculations, making it an ideal benchmark for validation.
Furthermore, the model's dimensionless formulation ensures scale invariance. This property allows the model to generalize to other astrophysical systems, such as supermassive black hole lenses in the LISA band.

The resulting emulator achieves sub-percent accuracy across the training domain while preserving high-frequency interference features, yielding large speedups relative to direct evaluations.
We take the PML as our baseline model, but we also extend the approach to more realistic lenses such as the SIS.
In addition, we discuss the astrophysical relevance of the parameter ranges, including stellar-mass lenses and IMBH or halo substructure in space-based observations \citep{pagano2020lensinggw,tambalo2023gravitational,wright2022gravelamps}.
The implementation is available at Github \footnote{\url{https://github.com/zqk-7k/Point-mass-lens-amplification-simulation}}.

This paper is organized as follows: 
In Sec.~\ref{sec:Background}, we review the theoretical framework of wave-optics lensing for the PML model.
Sec.~\ref{sec:num_methods} provides an overview of traditional numerical integration techniques and introduces our proposed SIREN-based architecture, supported by a theoretical analysis of its spectral properties.
Sec.~\ref{sec:implementation} outlines the experimental setup, detailing the data generation process and our physics-informed domain decomposition strategy.
Sec.~\ref{sec:results} presents an indepth performance evaluation, including accuracy benchmarks, computational efficiency comparisons against asymptotic expansion methods, and a validation of the model's generalization capability on the SIS profile.
Finally, Sec.~\ref{sec:conclusions} summarizes our findings and discusses future applications.
In this work, we adopt the $\Lambda$CDM cosmology model described in ~\cite{2020A&A...641A...6P}, which is also coded in \textbf{Astropy}\footnote{http://www.astropy.org} \citep{2022ApJ...935..167A} with assumptions of  $H_{0}=67.66\mathrm{km}\cdot\mathrm{s}^{-1}\cdot\mathrm{Mpc}^{-1},\Omega_{\mathrm{m}}=0.30966, T_{\mathrm{CMB}}=2.7255 \mathrm{K}$.

\section{Theoretical Background: Gravitational lensing of PML} \label{sec:Background}

As motivated in the Introduction, we adopt the PML as the baseline for CBCs in the LVK band, and work with the complex amplification factor \(F(\omega,y)\).

\subsection{Diffraction Integral Formulation}

\begin{figure}[htbp]
    \centering
    \includegraphics[width=\columnwidth]{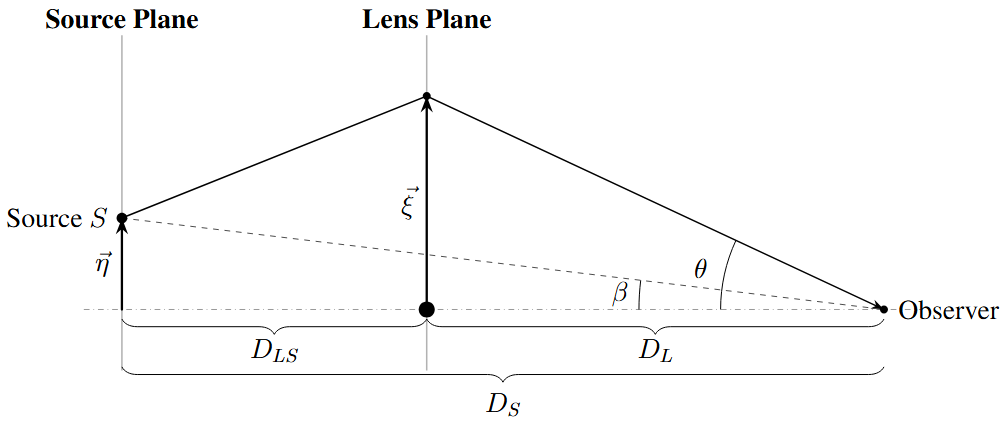}
    \caption{Schematic illustration of the thin-lens gravitational lensing geometry. The true source angular position ${\beta}$ corresponds to vector $\vec{{\eta}}$ in the source plane, while the observed angular position ${\theta}$ corresponds to the impact vector $\vec{{\xi}}$ in the lens plane. $D_L$, $D_S$, and $D_{LS}$ denote the relevant angular diameter distances to the lens and source planes.}
    \label{fig:lensing_geometry}
\end{figure}

We adopt the standard geometric configuration for gravitational lensing under the thin-lens approximation, as illustrated schematically in Fig.~\ref{fig:lensing_geometry}. A GW signal emitted by a source $S$ at a true angular position ${\beta}$ is deflected by a lens mass before reaching the observer. Due to this deflection, the observer perceives the signal coming from an apparent angular position ${\theta}$. As shown in Fig.~\ref{fig:lensing_geometry}, these angular positions correspond to physical coordinate vectors $\vec{{\eta}}$ in the source plane and $\vec{{\xi}}$ (the impact parameter) in the lens plane, respectively. The system is characterized by the angular diameter distances from the observer to the lens $D_L$, to the source $D_S$, and between the lens and the source $D_{LS}$.

Following~\cite{christian2018detecting}, we define the normalized coordinates in the lens plane $\vec{x}$, source plane $\vec{y}$ and observer plane $\vec{d}$ by the Einstein radius $\xi_0$:
\begin{equation}
\vec{x} = \frac{\vec{\xi}}{\xi_0}, \quad 
\vec{y} = \frac{D_L}{D_S} \frac{\vec{\eta}}{\xi_0}, \quad 
% \vec{d} = \left(1 - \frac{D_L}{D_S} \right) \frac{\vec{\Delta}}{\xi_0},
\end{equation}
where $\vec{\xi}$, $\vec{\eta}$, and $\vec{\Delta}$ represent the physical position vectors in the lens, source, and observer planes, respectively. The Einstein radius $\xi_0$ serves as the natural length scale.

The amplification factor $F(\omega, y)$, defined as the ratio of the lensed and unlensed GW amplitudes, is computed by solving the Fresnel–Kirchhoff diffraction integral in the thin-lens approximation:
\begin{equation} \label{eq:diffraction_integral}
F(\omega, y) = \frac{\omega}{2\pi i} \int \mathrm{d}^2 \vec{x} \, \exp\left[i \omega T(\vec{x}, \vec{y})\right],
\end{equation}
where $T(\vec{x}, \vec{y})$ is the time delay function.
% which includes both the geometric ($\vec{x}$, $\vec{y}$, and $\vec{d}$) and gravitational (lens potential $ \psi(\vec{x})$) contributions.
% , is given by:
% \begin{equation}
% T(\vec{x}, \vec{y}) = \frac{1}{2} (\vec{x} - \vec{y} - \vec{d})^2 - \psi(\vec{x}).
% \end{equation}
And the dimensionless frequency $\omega$, related to the physical frequency $f$, is given by:
\begin{equation} \label{eq:omega}
\omega = \frac{D_S \xi_0^2}{D_L D_{LS}} \tilde{\omega},
\end{equation}
where $\tilde{\omega} = 2\pi f$ is the angular frequency of the GW.

\subsection{Amplification Factor for PML}

The amplification factor for a PML can be derived by solving the diffraction integral, leading to the following closed-form expression for $F(\omega, y)$ ~\citep{takahashi2003wave,christian2018detecting}:

\begin{align} \label{eq:Fomega}
F(\omega, y) =\; & \exp\left\{ \frac{\pi \omega}{4} + i\frac{\omega}{2} \left[ \log\left( \frac{\omega}{2} \right) - 2\phi_m(y) \right] \right\} \notag \\
& \times \Gamma\left( 1 - \frac{i\omega}{2} \right)
{}_1F_1\left( \frac{i\omega}{2},\, 1;\, \frac{i\omega y^2}{2} \right),
\end{align}
where ${}_1F_1(a,b,z)$ is the confluent hypergeometric function of the first kind and $\Gamma$ is the Gamma function. The term $\phi_m(y)$ represents the minimum time delay and is defined by:
\begin{equation}
\phi_m(y) = \frac{1}{2}(x_m - y)^2 - \log x_m,  \quad x_m = \frac{y + \sqrt{y^2 + 4}}{2},
\end{equation}
here $x_m$ is the position in the lens plane where the time delay is minimized, and $y$ is the dimensionless source-lens separation in units of the Einstein radius.

For a PML, $\omega$ in Eq.~\eqref{eq:omega}  can be written as:
\begin{equation} \label{eq:omega_pm}
\omega = \frac{8\pi G M_L (1+z_L)}{c^3}f,
\end{equation}
where $M_L$ is the mass of point lens and $z_L$ is the lens redshift. This frequency is widely used as the standard parameter to distinguish the wave-optics and geometric-optics regimes. If $\omega \gg 1$, under this situation the geometric-optics approximation is applicable. If the dimensionless frequency $w$ does not satisfy this condition, the full diffraction integral incorporating wave-optics effects should be evaluated. Stellar mass black holes can serve as representative lens objects in the PML model. For a lens mass of $M_L = 60\,M_\odot$ at a redshift $z_L = 1$, and considering the most sensitive frequency band ($10$--$1000\,\mathrm{Hz}$) of current ground-based GW detectors, the corresponding dimensionless frequency range computed using Eq.~\eqref{eq:omega_pm} is $w\in[0.15,\,14.85]$. In this regime, the geometric-optics approximation may not be sufficient. For PML with smaller masses or located at lower redshifts, the upper bound of $w$ becomes even smaller, further emphasizing the necessity of computing the full amplification factor that includes wave-optics effects.

These expressions provides a detailed description of the amplification factor for lensed GWs and serves as the reference model for this study. The complex nature of $F(\omega, y)$, particularly its highly oscillatory behavior at large $y$, highlights the need for accurate numerical methods or machine learning-based approximations to efficiently model this function.

\subsection{Physical Interpretation and Illustrative Behavior}

The amplification factor \(F(\omega,y)\) depends critically on two dimensionless inputs: frequency \(\omega\) and impact parameter \(y\). In the low-frequency limit (\(\omega < 1\)), diffraction effects dominate and \(F\) approaches unity with a smooth envelope. As \(\omega\) increases, interference fringes appear and become more closely spaced, reflecting rapid phase accumulation between multiple propagation paths. Similarly, small \(y\) values (near alignment) produce strong magnification peaks at low frequency, while larger \(y\) values weaken the overall amplitude but generate high-frequency, dense oscillations as path-length differences grow.

\begin{figure*}[t]
  \centering
  \includegraphics[width=\textwidth]{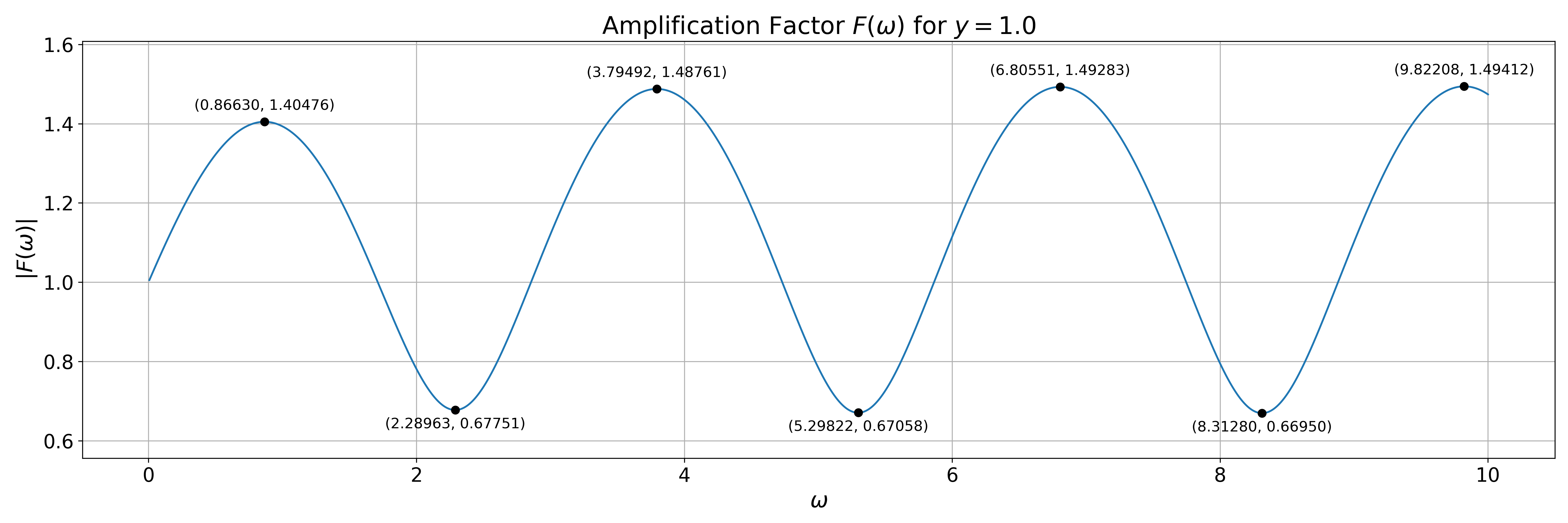}
  \caption{Illustration of \(\lvert F(\omega,y) \rvert\) for \(\omega \in [0.006, 10]\) at \(y=1.0\). This impact parameter corresponds to the Einstein radius, serving as a representative intermediate position where interference fringes are clearly visible. For a fiducial lens configuration with \(M_L=30\,M_\odot\) and \(z_L=0.25\), this \(\omega\) range maps to a physical band of \(1.4~\mathrm{Hz}\) to \(2.2~\mathrm{kHz}\), effectively covering the \(10\text{--}1000~\mathrm{Hz}\) sensitivity window of ground-based GW detectors. The non-periodic, sharply pointed oscillations highlight the necessity for a model capable of capturing multi-scale interference patterns.}
  \label{fig:F_example}
\end{figure*}

To visualize the characteristic behavior of the amplification factor in the wave-optics regime, we present a representative case in Fig.~\ref{fig:F_example} with a dimensionless impact parameter y=1.0. In this figure, interference fringes are clearly resolved.
We fix the lens redshift at \(z_L=0.25\) and consider a lens mass of \(M_L=30\,M_\odot\). According to Eq.~\eqref{eq:omega_pm}, the dimensionless frequency range will be \(\omega \in [0.006, 10]\) maps to a physical frequency band of \(f = [1.4~\mathrm{Hz}, 2.2~\mathrm{kHz}]\), which effectively covering the sensitivity window of ground-based GW detectors.
As illustrated in Fig.~\ref{fig:F_example}, the resulting amplification factor \(|F(\omega,y)|\) exhibits highly non-stationary behavior: the oscillation frequency increases with \(\omega\), reflecting the rapid phase accumulation between interfering paths, while the amplitude modulation follows a slow envelope.
The pattern is neither strictly periodic nor purely sinusoidal, and each extremum features a sharp peak.
This multi-scale structure, which combines a broad diffractive envelope with dense, high-frequency fringes, poses a severe challenge for low-order basis expansions or naive interpolation.
This complexity motivates our adoption of a learning model capable of capturing both global trends and fine-grained spectral features.

\section{Numerical Methods Overview and model}
\label{sec:num_methods}

% --- 将原本的 3.1.6 移到这里，作为全章的 Motivation ---

In this section, we first review several established numerical techniques (Sec.~\ref{sec:traditional_methods}) to establish a baseline, then introduce our SIREN–based model (Sec.~\ref{sec:siren_model}) that leverages learned high‐frequency bases to achieve sub‐percent accuracy with dramatically reduced run‐time.

\subsection{Traditional Numerical Methods}
\label{sec:traditional_methods}

Following standard wave–optics treatments \citep{Guo2020}, the amplification factor for an axisymmetric PML can be written as:
\begin{equation}
\begin{aligned}
F(\omega,y)
&= \frac{\omega}{i}\,
   \exp\!\Bigl[i\,\omega\Bigl(\tfrac{y^2}{2}+\phi_m(y)\Bigr)\Bigr] \\
&\quad\times \int_{0}^{\infty} x\,\mathrm{d}x
      \exp\!\Bigl[i\,\omega\Bigl(\tfrac{x^2}{2}-\psi(x)\Bigr)\Bigr]\,
      J_{0}\!\bigl(\omega x y\bigr)\,
\end{aligned}
\label{eq:amplification}
\end{equation}

where $J_0$ is the Bessel function of the first kind.
For the definitions of other variables including the dimensionless frequency $\omega$, impact parameter $y$, and the lens potential, we refer the reader to Sec.~\ref{sec:Background}. Substituting $z=x^2/2$, the diffraction integral transforms to:
% where $\omega$ is the dimensionless frequency, $y$ is the impact parameter, $\psi(x)=\ln x$ is the point–mass lens potential (up to an additive constant), $J_0$ is the Bessel function of the first kind, and
% \[
%   \phi_m(y)=\tfrac{1}{2}\bigl(x_m-y\bigr)^2-\ln x_m,\qquad
%   x_m=\tfrac{1}{2}\!\left(y+\sqrt{y^2+4}\right)
% \]
% is the minimum–time–delay phase. Introducing the change of variables $z=x^2/2$ (so that $x\,\mathrm{d}x=\mathrm{d}z$), the core diffraction integral is equivalently
\begin{equation}
  I(\omega,y)\;=\;\int_{0}^{\infty}
    \exp\!\bigl[i\,\omega\bigl(z-\psi(\sqrt{2z})\bigr)\bigr]\,
    J_{0}\!\bigl(\omega\,y\,\sqrt{2z}\bigr)\,\mathrm{d}z,
  \label{eq:diffraction_integral}
\end{equation}
and $F(\omega,y)=\frac{\omega}{i}\,e^{i\omega(y^2/2+\phi_m(y))}\,I(\omega,y)$.

\subsubsection{Integral Mean Method}
We approximate the improper integral \eqref{eq:diffraction_integral} by its $(C,1)$ Ces\`aro mean:
\begin{equation}
  I_C(b)
  \;=\;\int_{0}^{b}\!\Bigl(1-\frac{z}{b}\Bigr)\,
    \exp\!\bigl[i\,\omega\bigl(z-\psi(\sqrt{2z})\bigr)\bigr]\,
    J_{0}\!\bigl(\omega y\sqrt{2z}\bigr)\,\mathrm{d}z,
  \label{eq:integral_mean}
\end{equation}
which damps late–time oscillations and converges in the mean as $b\to\infty$.

\subsubsection{Asymptotic Expansion Method}
Split the domain at a finite cutoff \(b\) and write the integrand as \(e^{i \omega z} f(z)\), where

\begin{equation}
  f(z)=\exp\!\bigl[-\,i \omega\,\psi(\sqrt{2z})\bigr]\,
        J_{0}\!\bigl(\omega\,y\,\sqrt{2z}\bigr).
\end{equation}
Repeated integration by parts on the tail \([b,\infty)\) leads to the asymptotic–expansion
estimator
\begin{equation}
  I_{\mathrm{AE}}(b)
  \;=\; \int_{0}^{b} e^{i \omega z}\,f(z)\,\mathrm{d}z
  \;+\; e^{i \omega b}\sum_{n=1}^{N}\frac{(-1)^n}{(i \omega)^{n}}\,f^{(n-1)}(b)\,,
  \label{eq:asymp_expansion}
\end{equation}
where \(N\) is the truncation order and \(f^{(n-1)}(b)\) denotes the \((n\!-\!1)\)th derivative of \(f\) at \(z=b\).

\subsubsection{Levin’s Method}
Levin’s method computes oscillatory integrals on a finite interval,
\begin{equation}
  I(b) \;=\; \int_{a}^{b} f(z)\,e^{iq(z)}\,\mathrm{d}z,
  \label{eq:levin_setup}
\end{equation}
where $f(z)$ is a nonoscillatory amplitude and $q(z)$ is a smooth phase. 
Choose $n$ smooth, linearly independent basis functions $\{u_k(z)\}_{k=1}^n$ and
collocation nodes $\{z_j\}_{j=1}^n\subset[a,b]$. The coefficients 
$\{\alpha_k\}_{k=1}^n$ are determined by the collocation system
\begin{equation}
\begin{aligned}
\sum_{k=1}^{n}\alpha_k\,u_k'(z_j)
&\;+\; i\,q'(z_j)\sum_{k=1}^{n}\alpha_k\,u_k(z_j)
= f(z_j),\\
&\qquad j=1,\ldots,n.
\end{aligned}
\label{eq:levin_collocation}
\end{equation}

The collocated antiderivative yields the endpoint expression
\begin{equation}
  I_n(b)\;=\;\sum_{k=1}^{n}\alpha_k\,u_k(b)\,e^{\,i\,q(b)}
  \;-\;\sum_{k=1}^{n}\alpha_k\,u_k(a)\,e^{\,i\,q(a)}.
  \label{eq:levin_Inb}
\end{equation}
We approximate the finite-interval integral by \(I(b)\approx I_n(b)\).
In our setting the integration variable is \(z=x^2/2\) with domain \([0,\infty)\), so we set
\(a=0\) and \(q(z)=\omega z\). Sending \(b\to\infty\) and assuming the endpoint term at \(b\) vanishes,
we obtain
\begin{equation}
  I(\infty) \;\approx\; -\sum_{k=1}^{n}\alpha_k\,u_k(0).
  \label{eq:levin_Iinfty}
\end{equation}

\subsubsection{Zero Points Integral Method}
Define the partial integral
\begin{equation}
  I(b)\;\equiv\;\int_{0}^{b}
  e^{\,i \omega\,(z-\psi(\sqrt{2z}))}\,
  J_{0}\!\bigl(\omega\,y\,\sqrt{2z}\bigr)\,\mathrm{d}z.
  \label{eq:I_b}
\end{equation}

Let $j_k$ be the $k$th positive root of $J_0$. Under $z=x^2/2$,

\begin{equation}
  z_k = \frac{j_k^{2}}{2\,\omega^{2}\,y^{2}},
  \label{eq:zp_position_z}
\end{equation}
\begin{equation}
  J_0\!\bigl(\omega\,y\,\sqrt{2z_k}\bigr) = 0.
  \label{eq:zp_position_J0}
\end{equation}
Evaluating \eqref{eq:I_b} at $b=z_k$ gives
\begin{equation}
  I(z_k)\;=\;\int_{0}^{z_k}
  e^{\,i \omega\,(z-\psi(\sqrt{2z}))}\,
  J_0\!\bigl(\omega\,y\,\sqrt{2z}\bigr)\,\mathrm{d}z.
  \label{eq:I_zk}
\end{equation}
The improper integral is
\begin{equation}
  I(\omega,y)\;=\;\lim_{k\to\infty} I(z_k).
  \label{eq:I_def}
\end{equation}
A $(C,1)$ Ces\`aro average often accelerates convergence:
\begin{equation}
  I_{\mathrm{CS}}(\infty)\;=\;\lim_{m\to\infty}\,\frac{1}{m}\sum_{k=1}^{m} I(z_k).
  \label{eq:ICS}
\end{equation}

\subsubsection{Zero Points Asymptotic Expansion Method}\label{sssec:zpae}
With the zero points $z_k$ from Eq.~\eqref{eq:zp_position_z} and the partial integral $I(z_k)$ from Eq.~\eqref{eq:I_zk}, the zero–points asymptotic expansion is
\begin{equation}
  I(\infty)\;=\;I(z_k)\;+\;e^{\,i \omega z_k}\,
  \sum_{n=1}^{\infty}\frac{(-1)^n}{(i \omega)^{n}}\,
  \frac{\partial^{\,n-1} f}{\partial z^{\,n-1}}\bigg|_{z=z_k}.
  \label{eq:zp_total}
\end{equation}
For accelerated convergence, one may apply a $(C,1)$ Ces\`aro average to the
asymptotic–expansion estimator \(I_{\mathrm{AE}}(b)\) in Eq.~\eqref{eq:asymp_expansion} evaluated at $b=z_k$:
\begin{equation}
  I_{\mathrm{CS,AE}}(\infty)\;=\;\lim_{m\to\infty}\frac{1}{m}
  \sum_{k=1}^{m} I_{\mathrm{AE}}(z_k).
  \label{eq:zp_total_csae} % 建议添加 label 以便引用
\end{equation}

\subsection{SIREN-based Neural Network Model}
\label{sec:siren_model}

\begin{figure}[tb]
  \centering
  \includegraphics[width=\columnwidth]{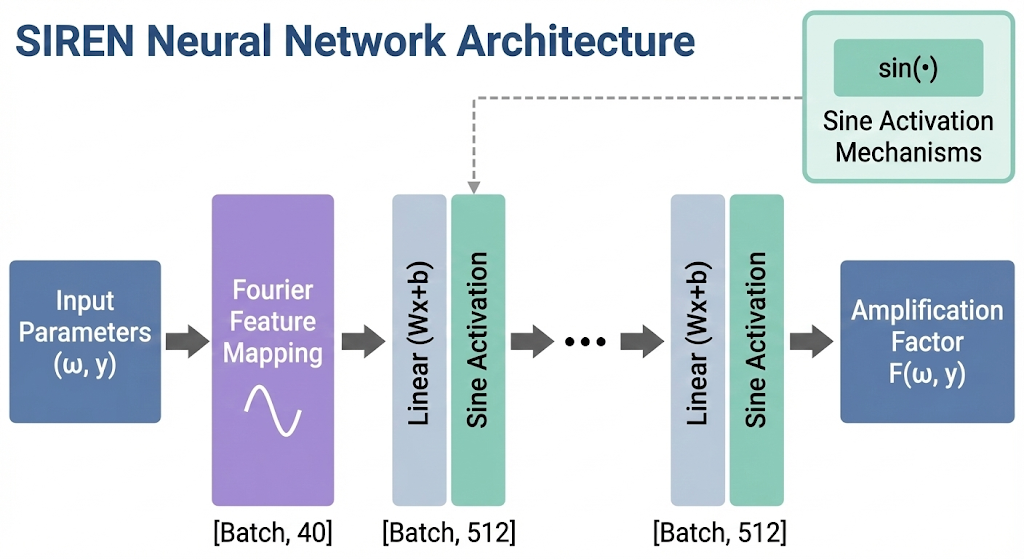}
    \caption{Architecture of the proposed neural model. The continuous input coordinates, representing the dimensionless gravitational wave frequency \(\omega\) and the source-lens impact parameter \(y\), are first projected onto a high-dimensional manifold via a fixed Gaussian Fourier feature mapping. The embedded features are then processed by a deep SIREN, where each layer employs a periodic sine activation function to naturally capture the multi-scale oscillatory behavior of the amplification factor.}
  \label{fig:siren_architecture}
\end{figure}

The wave-optics amplification factor $F(\omega, y)$ is characterized by intense oscillatory behavior, where dense interference fringes modulate a diffractive envelope.
Numerically, direct evaluation of Eq.~\eqref{eq:Fomega} is hindered by high computational costs and instability arising from the cancellation of special function terms in destructive-interference intervals. 
From a deep learning perspective, standard MLPs are also ill-suited for this task as they suffer from spectral bias, a phenomenon where networks prioritize low-frequency features and fail to resolve rapid variations~\citep{rahaman2019spectral}. In fact, our empirical tests also confirmed that standard MLPs completely fail to handle this highly oscillatory task. To overcome this limitation, we implement a coordinate-based neural network that integrates Fourier feature mappings~\citep{tancik2020fourier} with SIRENs (see Fig.~\ref{fig:siren_architecture}).

\subsubsection{Fourier Feature Mapping}
Before entering the network, the low-dimensional input coordinates \(\mathbf{x} = (\omega, y)\) are projected into a higher-dimensional feature space using a Fourier feature mapping \(\gamma: \mathbb{R}^2 \to \mathbb{R}^{2m}\). This mapping enables the network to learn high-frequency functions more efficiently by controlling the spectral decay of the Neural Tangent Kernel (NTK)~\citep{tancik2020fourier}. Specifically, we employ a Gaussian random Fourier feature mapping defined as:
\begin{equation}
\gamma(\mathbf{x}) = \left[ \cos(2\pi \mathbf{B}\mathbf{x}),\, \sin(2\pi \mathbf{B}\mathbf{x}) \right]^\mathsf{T},
\end{equation}
where \(\mathbf{B} \in \mathbb{R}^{m \times 2}\) is a fixed weight matrix containing \(m\) sampled frequency vectors, with entries sampled from a Gaussian distribution \(\mathcal{N}(0, \sigma^2)\). This projection maps the input onto a periodic manifold, enriching the input representation with high-frequency components prior to the learnable layers.

\subsubsection{Sinusoidal Representation Network}
The embedded features \(\mathbf{z}_0 = \gamma(\mathbf{x})\) serve as the input to the SIREN backbone. Unlike standard MLPs, SIREN employs a periodic sine activation function \(\sigma(x) = \sin(\omega_0 x)\) for every hidden layer. The network operation is defined recursively as:
\begin{equation}
\mathbf{z}_{i} = \sin\left(\omega_0 (\mathbf{W}_i \mathbf{z}_{i-1} + \mathbf{b}_i)\right), \quad i=1,\dots,L-1,
\end{equation}
followed by a final linear transformation \(\mathbf{z}_{L} = \mathbf{W}_L \mathbf{z}_{L-1} + \mathbf{b}_L\) to output the real and imaginary parts of the amplification factor. Here, \(\omega_0\) is a hyperparameter that dictates the frequency of the activation function.

The periodic nature of the activation function provides a strong structural consistency suited for wave physics, allowing the network to represent derivatives and fine-grained structures accurately. To ensure stable signal propagation, we adopt the specialized initialization scheme proposed by ~\cite{sitzmann2020implicit}. For the first hidden layer (accepting the Fourier features), weights are initialized from a uniform distribution:
\begin{equation}
\mathbf{W}_1 \sim \mathcal{U}\left(-\frac{1}{d_{\text{in}}}, \frac{1}{d_{\text{in}}}\right),
\end{equation}
where \(d_{\text{in}}\) is the dimensionality of the input Fourier features (i.e., \(2m\)). For subsequent layers, weights are initialized to preserve the activation variance:
\begin{equation}
\mathbf{W}_i \sim \mathcal{U}\left(-\frac{\sqrt{6/n_{\text{h}}}}{\omega_0}, \frac{\sqrt{6/n_{\text{h}}}}{\omega_0}\right), \quad i > 1,
\end{equation}
where \(n_{\text{h}}\) denotes the number of neurons in the hidden layers. This initialization strategy prevents the gradients from vanishing or exploding, enabling the training of deep architectures capable of resolving the complex interference patterns of \(F(\omega, y)\).

\subsection{Theoretical Motivation of the Architecture}
\label{sec:theoretical_analysis}

\begin{figure*}[htbp]
    \centering
    \includegraphics[width=0.95\textwidth]{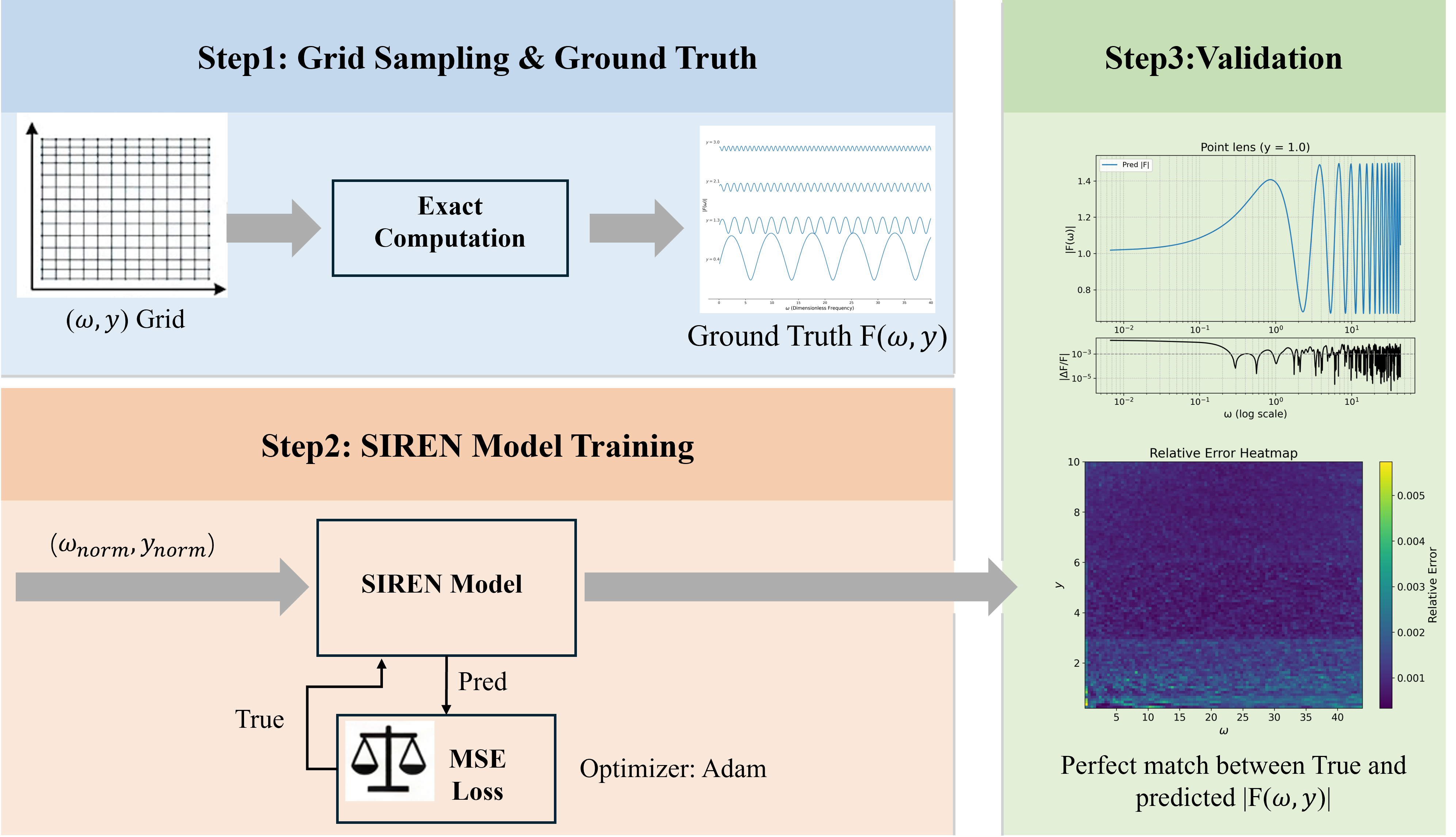} 
    \caption{Computational framework for diffraction integral evaluation. The workflow proceeds in three logical steps. First, \textbf{Data Generation} utilizes \texttt{mpmath} for high-precision analytical evaluation of the amplification factor \(F(\omega, y)\) on a dense grid. The selected grid astrophysically encompasses the transition from strong lensing to the weak-lensing diffraction tail, mapping universally to both stellar-mass lenses in ground-based detector bands and supermassive black holes in the LISA band. Second, \textbf{Model Training} employs a SIREN network enhanced with Fourier feature embeddings to learn the continuous mapping from normalized parameters to the complex amplification factor. Third, \textbf{Validation} compares the predicted spectral response \(\lvert F(\omega)\rvert\) against the ground truth to ensure accurate capture of high-frequency interference patterns.}
    \label{fig:simulation_pipeline}
\end{figure*}

We employ the NTK~\citep{jacot2018neural} to analyze the training dynamics, characterizing the convergence rate of neural networks in the frequency domain. First, consider the spectral properties of the target function. As shown in Eq.~\eqref{eq:amplification}, the amplification factor \(F(\omega, y)\) is composed of a rapidly oscillating exponential prefactor and a Bessel-weighted diffraction integral.
In the wave-optics regime (large \(\omega\) and \(y\)), the local oscillation frequency is determined by the gradient of the phase terms. Specifically, the characteristic bandwidth \(\nu_{\text{target}}\) is proportional to the product of the physical parameters:
\begin{equation}
    \nu_{\text{target}} \sim \omega y.
\end{equation}
This implies that the spectral support of \(F(\omega, y)\) extends to high frequencies, requiring the model to resolve fine-scale variations.

Next, consider the learning process. For a network \(f(\mathbf{x}; \theta)\) trained by gradient descent, the training dynamics in the infinite-width limit decouple into independent eigenmodes of the NTK. The error decay for a specific eigenmode is governed by its eigenvalue \(\lambda_\nu\):
\begin{equation}
    |\hat{\epsilon}_\nu(t)| \propto e^{-\eta \lambda_\nu t},
\end{equation}
where \(\hat{\epsilon}_\nu(t)\) is the error component corresponding to the spectral frequency \(\nu\) of the target function's decomposition (i.e., the spatial frequency in the parameter space \((\omega, y)\)), \(\eta\) is the learning rate, and \(\lambda_\nu\) determines the convergence rate.
Note that \(\nu\) here characterizes the rate of variation of the function \(F\) with respect to its inputs, distinct from the physical GW frequency \(\omega\) which serves as an input coordinate.

For standard MLPs with ReLU activation, ~\citep{rahaman2019spectral} demonstrated that the NTK eigenvalues decay rapidly according to a power law, typically \(\lambda_\nu^{\text{ReLU}} \propto \nu^{-2}\). This induces a "spectral bias": for the high-frequency modes present in the wave-optics regime (large \(\nu_{\text{target}}\)), the convergence speed \(\lambda_{\nu_{\text{target}}}\) becomes negligibly small. Consequently, ReLU networks act as low-pass filters, effectively smoothing out the essential interference fringes.

In contrast, SIREN employs periodic activations \(\sigma(x) = \sin(\omega_0 x)\). The NTK spectrum of SIREN is not concentrated at zero but exhibits tunable bandwidth controlled by the initialization frequency \(\omega_0\). By setting \(\omega_0\) to match the characteristic scale of the diffraction integral, we ensure that \(\lambda_{\nu}\) remains significant across the target spectrum.
Structurally, the composition of sinusoidal layers synthesizes a function:
\begin{equation}
    f_{\text{SIREN}}(\mathbf{x}) \sim \sum_{k} c_k \sin(\mathbf{w}_k \mathbf{x} + \phi_k).
\end{equation}
This form is structurally analogous to the oscillatory nature of Eq.~\eqref{eq:amplification}. The correspondence between the network's spectral properties and the physical model ensures the accurate representation of fine-grained diffraction patterns.

\section{Model Implementation and Experimental Setup}
\label{sec:implementation}

The simulation workflow integrates data generation and SIREN training, as illustrated in Figure~\ref{fig:simulation_pipeline}. This pipeline integrates the precise analytical generation of training data, the feature encoding strategy, the training of the SIREN, and the subsequent validation in the frequency domain.

\subsection{Parameter Range Selection}

The choice of the input domains for \(\omega\) and \(y\) is dictated by both the physical behavior of the amplification factor \(F(\omega,y)\) and numerical stability considerations.

% We select three representative masses of the PML \(M_L=\{10,30,50\}\,M_\odot\), which target the stellar–mass black hole population routinely inferred in LVK catalogs. 

We select three representative masses of the PML \(M_L=\{10,30,50\}\,M_\odot\), which target the stellar--mass black hole population routinely inferred in LVK catalogs. These values are used solely to determine the physically relevant boundary limits of the dimensionless frequency \(\omega\) in the LVK sensitivity band, rather than to fix the lens mass during training. Based on the optimal sensitivity band of current GW detectors (Advanced LIGO, Virgo, and KAGRA), we select a frequency range of \(f \in [1, 1000]\,\text{Hz}\), corresponding to the typical GW frequencies emitted by CBCs.

Throughout, we convert this physical range to the dimensionless frequency \(\omega\) using Eq.~\eqref{eq:omega_pm}. Taking the union of these resulting intervals yields our final training domain:
\[
\omega\in[\omega_{\min},\omega_{\max}],
\]
where \(\omega_{\min}=0.0067\) and \(\omega_{\max}=43.93\). The model is trained directly on \(F(\omega,y)\), so the dependence on \(M_L\) is fully encoded through \(\omega\), and no mass-specific training is required. This range is critical as it covers the transition from the diffraction-dominated regime to the dense oscillatory fringes of the wave-optics limit.

It is worth noting that due to the scale-invariance of the dimensionless formulation, this domain is astrophysically universal.
Focusing on the prime LISA frequency band of \(f \in [10^{-4}, 10^{-1}]\,\text{Hz}\)~\citep{amaro2017laser}, we consider the redshifted lens mass \(M_{Lz} = (1+z_L)M_L\) as the primary physical parameter.
In this context, our training domain of \(\omega \in [0.0067, 43.93]\) is astrophysically versatile, as it effectively encompasses the diffraction patterns produced by \(M_{Lz}\) ranging from \(\sim 10^3\,M_\odot\) to \(\sim 10^7\,M_\odot\).
Thus, our selected parameter space provides a rigorous testbed applicable to a wide range of supermassive black hole lensing scenarios.

For the impact parameter \(y\), we set:
\[
y\in[y_{\min},y_{\max}],
\]
with \(y_{\min}=0.2\) and \(y_{\max}=10.0\). The lower bound avoids numerical instability in the analytic formula as \(y\to 0\), while the upper bound extends to the weak-lensing regime where \(F \approx 1\). Our primary focus for detection prospects lies in the high-magnification region \(y \le 3\).

\subsection{Training Dataset Construction}
\label{sec:training-data}

The analytic amplification factor \(F(\omega,y)\) (Eq.~\ref{eq:Fomega}) was evaluated on a dense Cartesian grid covering the selected domain. Special‐function computations were carried out using the \texttt{mpmath} library with global precision set to 50 decimal places. The resulting ground‐truth labels were validated against adaptive quadrature of the Fresnel diffraction integral and geometric-optics limits, achieving agreement within \(10^{-15}\).

\subsection{Feature Encoding}
\label{sec:feature-encoding}

Prior to network ingestion, the input coordinates \((\omega, y)\) are first normalized to the range \([-1,1]\) via linear scaling:
\[
  \omega_{\mathrm{norm}} = \frac{2(\omega - \omega_{\min})}{\omega_{\max}-\omega_{\min}} - 1,\quad
  y_{\mathrm{norm}} = \frac{2(y - y_{\min})}{y_{\max}-y_{\min}} - 1.
\]
Let \(\mathbf{x} = (\omega_{\mathrm{norm}},\,y_{\mathrm{norm}})\). As detailed in the model architecture (Sec.~\ref{sec:siren_model}), we enrich the input representation using a Gaussian Fourier feature mapping. In this implementation, we set the mapping dimension to \(m=20\) and the Gaussian scale to \(\sigma=10.0\). The resulting projected features \(\Phi(\mathbf{x}) \in \mathbb{R}^{40}\) are concatenated with the normalized inputs \(\mathbf{x}\) to form the final input vector passed to the SIREN backbone.

\subsection{Domain Decomposition and Sampling Strategy}
\label{sec:domain_decomposition}

\begin{table}[ht]
  \centering
  \caption{Adaptive sampling configuration for each interval \(\mathcal{I}_i\). \(N_y^{(i)}\) and \(N_\omega^{(i)}\) denote grid points in \(y\) and \(\omega\).}
  \label{tab:sampling}
  \begin{tabular}{|c|c|c|c|c|}
    \hline
    Interval & \(y\)-range & \(N_y^{(i)}\) & \(N_\omega^{(i)}\) & Total samples \\
    \hline
    \(\mathcal{I}_1\) & \([0.2,\,1.0)\) & 200 & 5\,000 & \(1.0\times10^6\) \\
    \(\mathcal{I}_2\) & \([1.0,\,3.0)\) & 400 & 5\,000 & \(2.0\times10^6\) \\
    \(\mathcal{I}_3\) & \([3.0,\,6.0)\) & 600 & 10\,000 & \(6.0\times10^6\) \\
    \(\mathcal{I}_4\) & \([6.0,\,10.0]\) & 800 & 20\,000 & \(1.6\times10^7\) \\
    \hline
  \end{tabular}
\end{table}

\begin{table}[ht]
  \centering
  \caption{Hyperparameter search and final selections.}
  \label{tab:hyperparams}
  \resizebox{\columnwidth}{!}{%
  \begin{tabular}{lcc}
    \hline
    Hyperparameter                  & Search Range               & Selected Value \\
    \hline
    SIREN base frequency $\omega_0$ & $[1,\,100]$                & 80             \\
    Network depth                   & $[4,\,12]$                 & 8              \\
    Layer width                     & $[128,\,1024]$             & 512            \\
    Fourier feature count ($m$)     & $[0,\,50]$                 & 20             \\
    Batch size                      & $[256,\,8192]$             & 4096           \\
    Initial learning rate           & $[10^{-6},\,5\times10^{-7}]$ & Interval-dependent \\
    \hline
  \end{tabular}
  }
\end{table}

\begin{table*}[t]
\centering
\caption{Summary of the model’s relative error performance. Evaluated on random off-grid samples, reported metrics include mean, median, 95th/99th percentiles, and maximum error across different $y$-intervals.}
\label{tab:metrics}
\begin{tabular}{lccccc}
\hline
Metric                   & Overall                & $\mathcal I_1$ [0.2, 1.0]           & $\mathcal I_2$ [1.0, 3.0]           & $\mathcal I_3$ [3.0, 6.0]           & $\mathcal I_4$ [6.0, 10.0] \\
\hline
Mean Rel.\ Error         & $1.10\times10^{-3}$  & $1.83\times10^{-3}$      & $1.58\times10^{-3}$      & $8.27\times10^{-4}$      & $8.75\times10^{-4}$ \\
Median Rel.\ Error       & $8.30\times10^{-4}$  & $1.47\times10^{-3}$      & $1.22\times10^{-3}$      & $6.94\times10^{-4}$      & $7.28\times10^{-4}$ \\
95th Percentile Error    & $3.02\times10^{-3}$  & $4.73\times10^{-3}$      & $4.22\times10^{-3}$      & $2.05\times10^{-3}$      & $2.18\times10^{-3}$ \\
99th Percentile Error    & $4.94\times10^{-3}$  & $7.40\times10^{-3}$      & $6.26\times10^{-3}$      & $2.74\times10^{-3}$      & $2.99\times10^{-3}$ \\
Maximum Error            & $3.23\times10^{-2}$  & $2.05\times10^{-2}$      & $3.23\times10^{-2}$      & $4.84\times10^{-3}$      & $1.26\times10^{-2}$ \\
\hline
\end{tabular}
\end{table*}

\begin{figure*}[ht]
  \centering
  \begin{subfigure}[t]{0.48\textwidth}
    \centering
    \includegraphics[width=\linewidth]{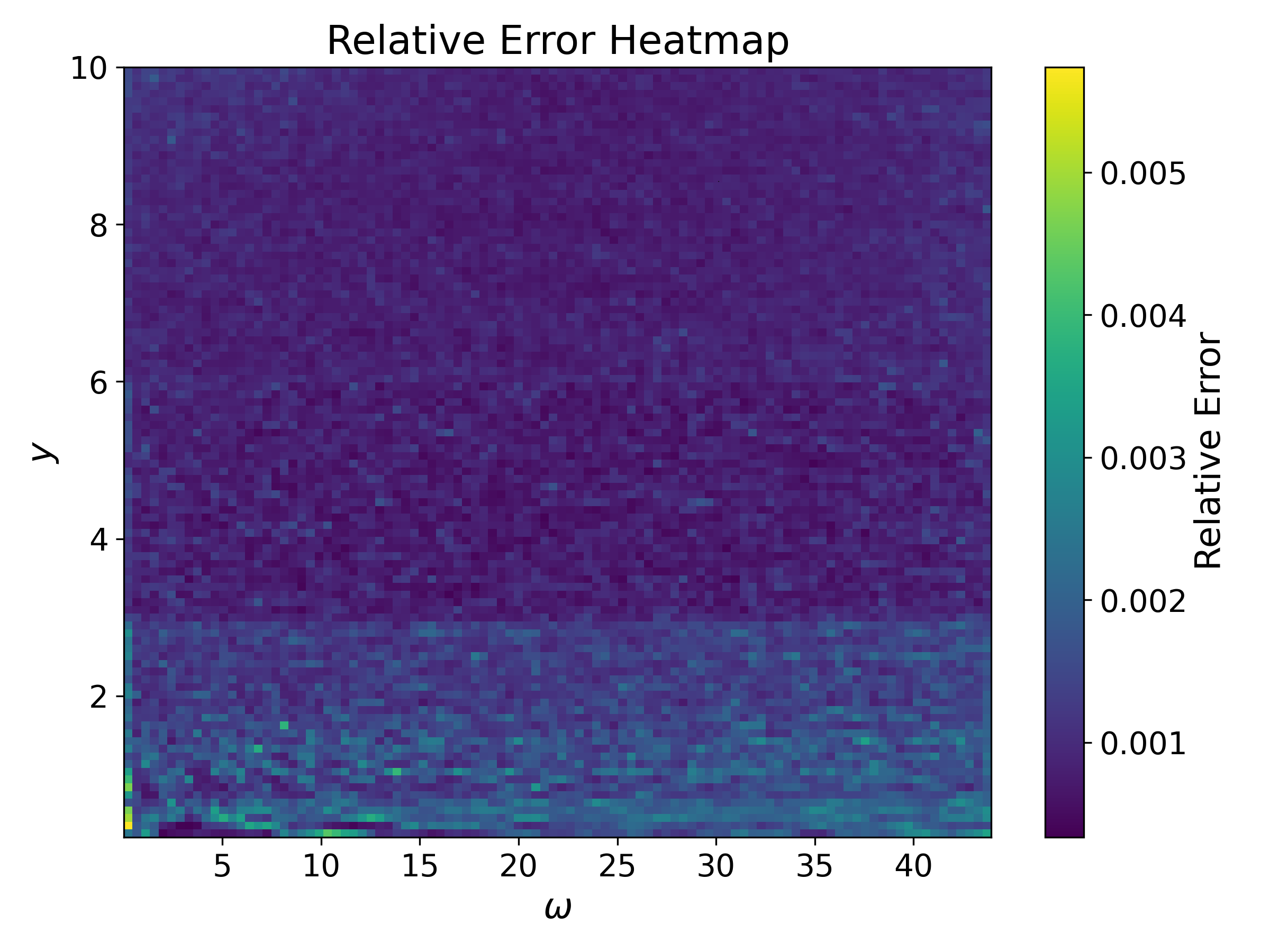}
    \caption{Training-grid samples}
    \label{fig:heatmap_train}
  \end{subfigure}
  \hfill
  \begin{subfigure}[t]{0.48\textwidth}
    \centering
    \includegraphics[width=\linewidth]{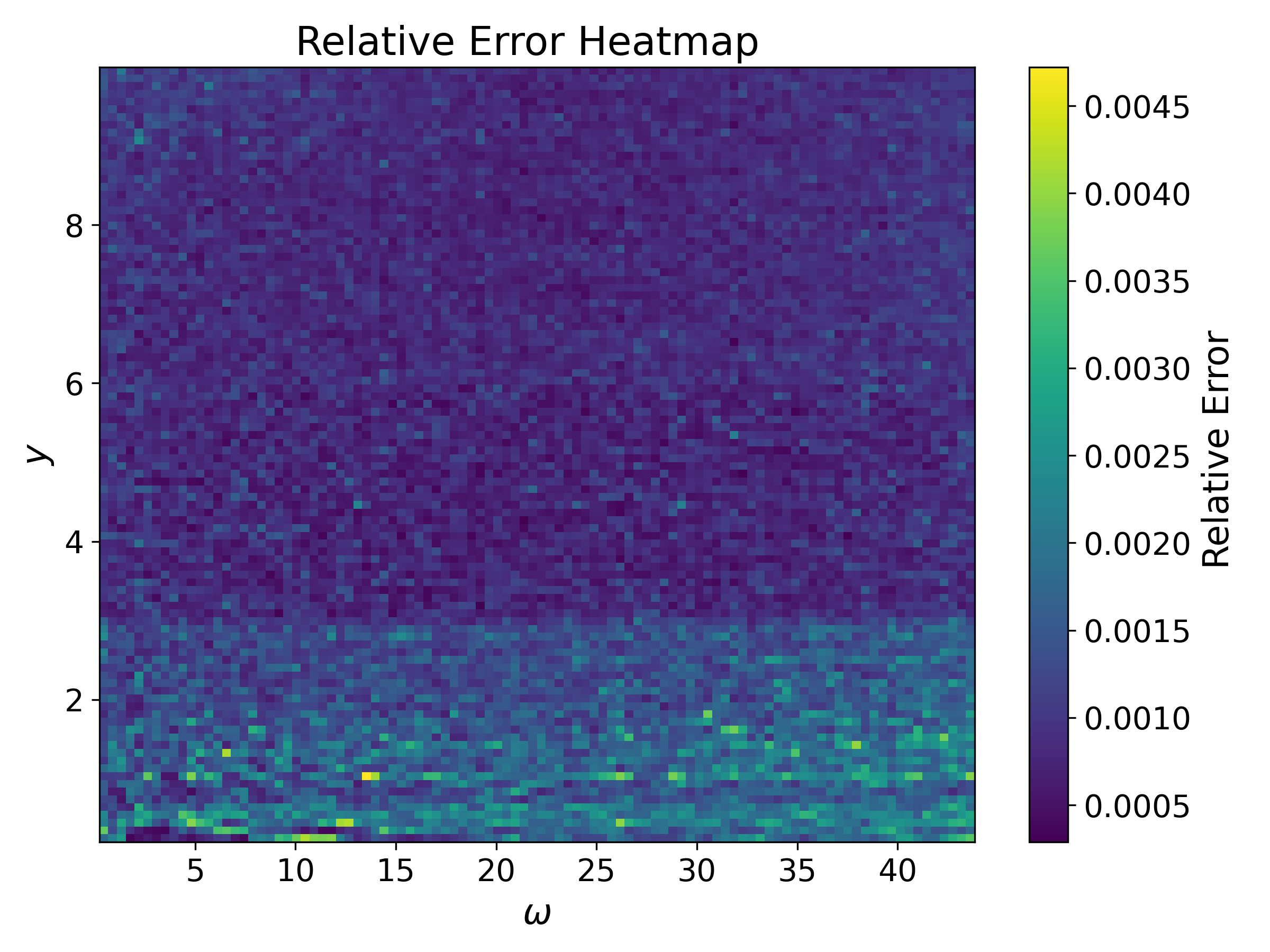}
    \caption{Random samples}
    \label{fig:heatmap_rand}
  \end{subfigure}
  \caption{Relative-error heatmaps showing $\epsilon_{\mathrm{rel}}$ as a function of dimensionless frequency $\omega$ and source position $y$.}
  \label{fig:error_heatmaps}
\end{figure*}

\begin{figure}[h!]
    \centering
    \includegraphics[width=0.48\textwidth]{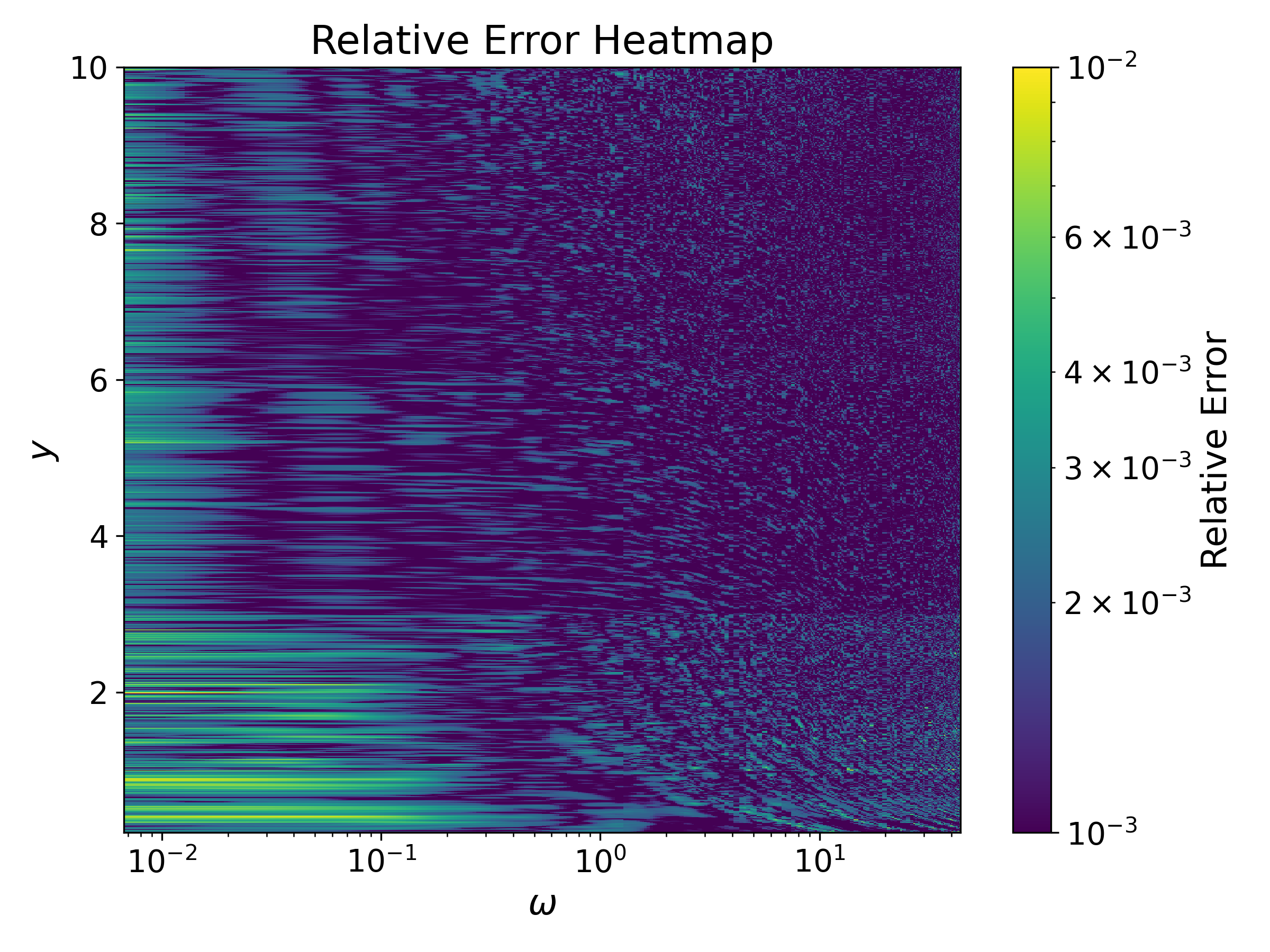}
    \caption{Heatmap of the relative error distribution across the parameter space. The transition from diffraction to interference ($y \sim 1.0$) shows high stability with errors well below the $\mathcal{O}(10^{-3})$ threshold.}
    \label{fig:response_heatmap}
\end{figure}

The amplification factor exhibits highly non-stationary behavior: smooth variations at low \(y\) and rapid, dense oscillations at high \(y\). This behavior poses a multi-scale challenge for unified modeling. Inspired by the ``Frequency Principle''~\citep{xu2019frequency}, which states that neural networks tend to fit low-frequency components first, we adopt a domain decomposition strategy to reduce the spectral dynamic range of each sub-task. Analogous to the adaptive integration intervals employed by ~\cite{Guo2020} for numerical integration, we decompose the \(y\)-domain into four overlapping intervals \(\mathcal{I}_i = [\alpha_i, \alpha_{i+1}]\) defined by breakpoints \(\{0.2, 1.0, 3.0, 6.0, 10.0\}\).

We employ a domain decomposition based on the interference characteristics: \(\mathcal{I}_1\) and \(\mathcal{I}_2\) cover the strong-lensing and high-contrast fringe regimes relevant to CBCs, while \(\mathcal{I}_3\) and \(\mathcal{I}_4\) handle the high-frequency saturation and weak-lensing tails. This allows us to tune the network bandwidth according to local oscillatory characteristics, preventing optimization stagnation common in single networks facing mixed-frequency tasks.

Within each interval, we employ an adaptive tensor-product grid for sampling. High-frequency intervals utilize finer grids (up to \(N_\omega=20,000\) points) to resolve dense fringes, while smoother regions use coarser sampling (see Table~\ref{tab:sampling}).

This modular design extends beyond training to the inference phase. By establishing a direct mapping in the parameter space \((\omega, y)\), the system automatically routes any query to the appropriate SIREN sub-network based on the input value of \(y\). Unlike traditional numerical integration, where cost scales with oscillation frequency due to sampling requirements, this routing mechanism decouples inference time from \(\omega\). This theoretically ensures a constant \(\mathcal{O}(1)\) computational complexity, a property we empirically verify in Section~\ref{sssec:efficiency}.

\subsection{Network Training Details}
\label{sec:training_details}

All models are implemented in PyTorch~2.0 and trained on NVIDIA A100 GPUs using mixed‐precision. We use the Adam optimizer with a plateau‐based scheduler. Separate loss functions are minimized for the real and imaginary components of the amplification factor.

To ensure numerical continuity across the domain decomposition boundaries (i.e., at \(y=1.0, 3.0, 6.0\)), we employed an overlapping training strategy.
Specifically, the boundary values are explicitly included in the training datasets of both adjacent intervals (e.g., \(y=1.0\) is present in both \(\mathcal{I}_1\) and \(\mathcal{I}_2\)).
Since both sub-models converge to the same high-precision analytical ground truth at these interfaces, the discrepancy between models at the boundaries is minimized to the order of the intrinsic validation error (\(\sim 10^{-3}\)).
This ensures that the piecewise inference remains sufficiently smooth for likelihood-based sampling methods, preventing artificial jumps in the parameter estimation posterior.
Key hyperparameters, selected via grid search, are summarized in Table~\ref{tab:hyperparams}.

\begin{figure*}[ht]
  \centering
  \begin{subfigure}[t]{\textwidth}
    \centering
    \includegraphics[width=0.8\textwidth]{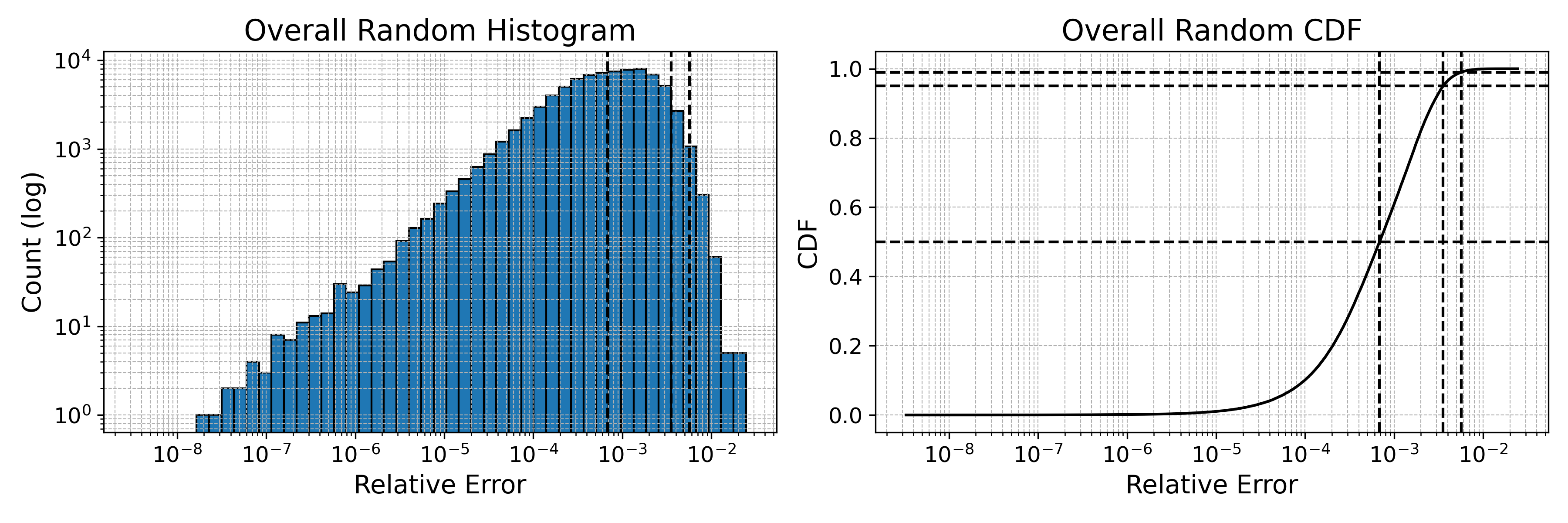}
    \caption{Random test points: histogram and CDF.}
    \label{fig:hist_rand}
  \end{subfigure}\\[1ex]
  \begin{subfigure}[t]{\textwidth}
    \centering
    \includegraphics[width=0.8\textwidth]{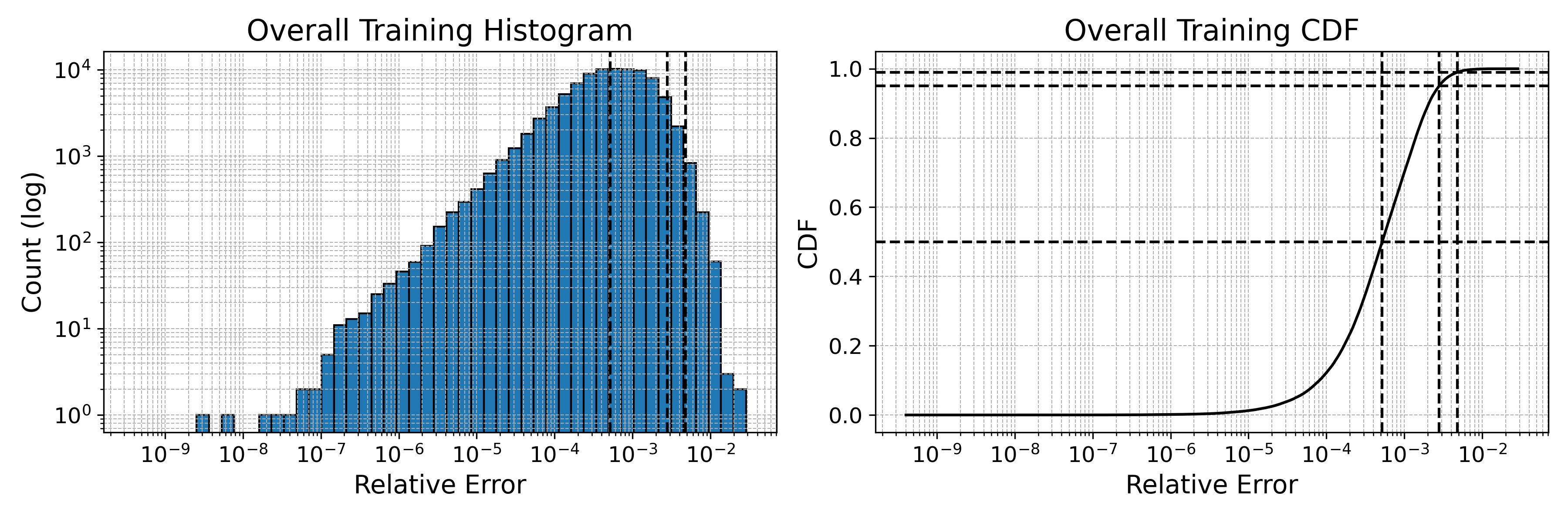}
    \caption{Training-grid points: histogram and CDF.}
    \label{fig:hist_train}
  \end{subfigure} 
  \caption{Overall distributions of relative error $\epsilon_{\rm rel}$. The nearly identical shape between (a) and (b) indicates excellent generalization.}
  \label{fig:hist_errors}
\end{figure*}

\begin{figure*}[htbp]
  \centering
  \begin{subfigure}[t]{0.48\textwidth}
    \centering
    \includegraphics[width=\linewidth]{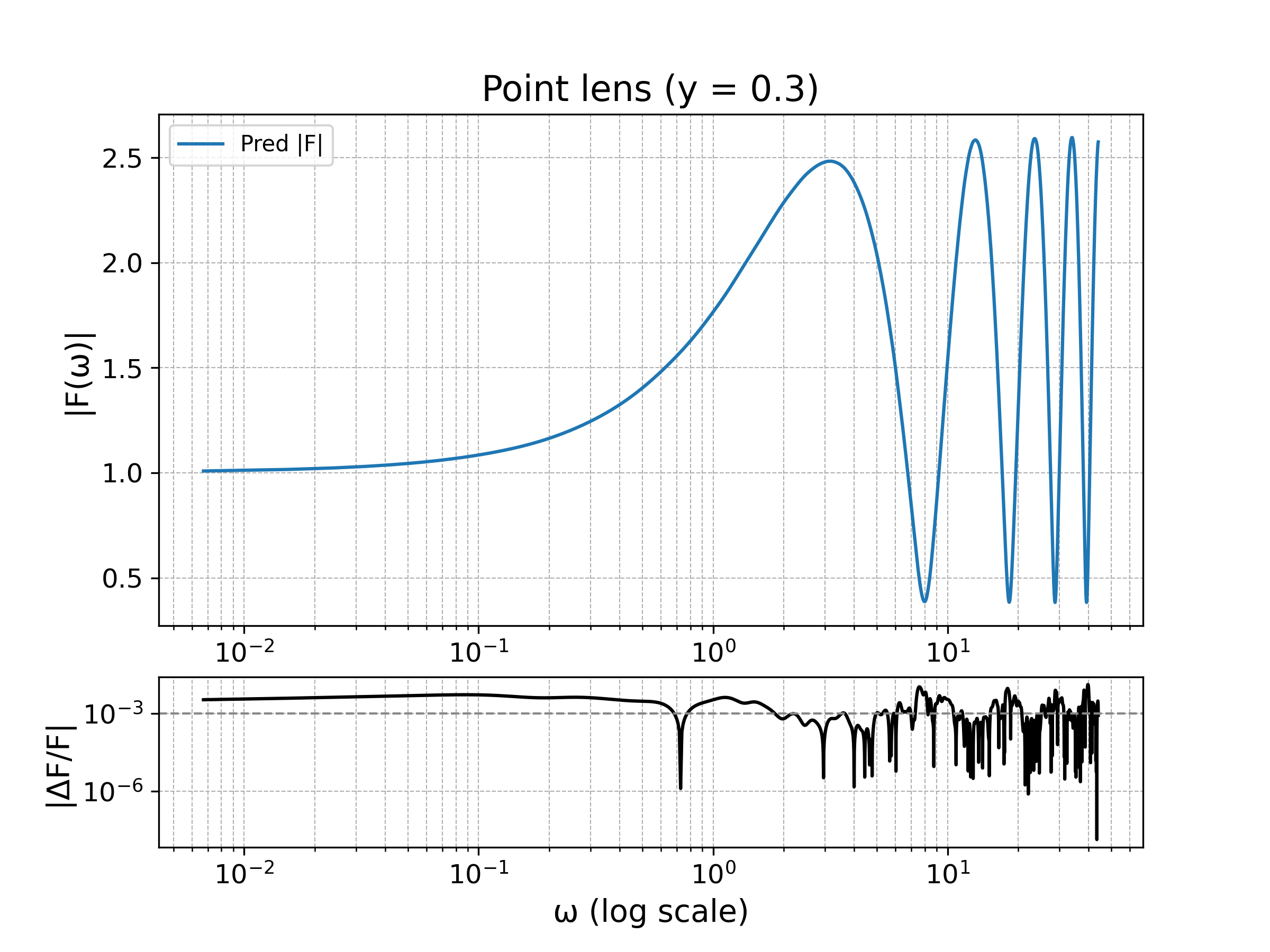}
    \caption{$y=0.3$: Pronounced principal diffraction peak.}
    \label{fig:pred_y0.3}
  \end{subfigure}
  \hfill
  \begin{subfigure}[t]{0.48\textwidth}
    \centering
    % Please ensure the filename extension matches your file (png/jpg)
    \includegraphics[width=\linewidth]{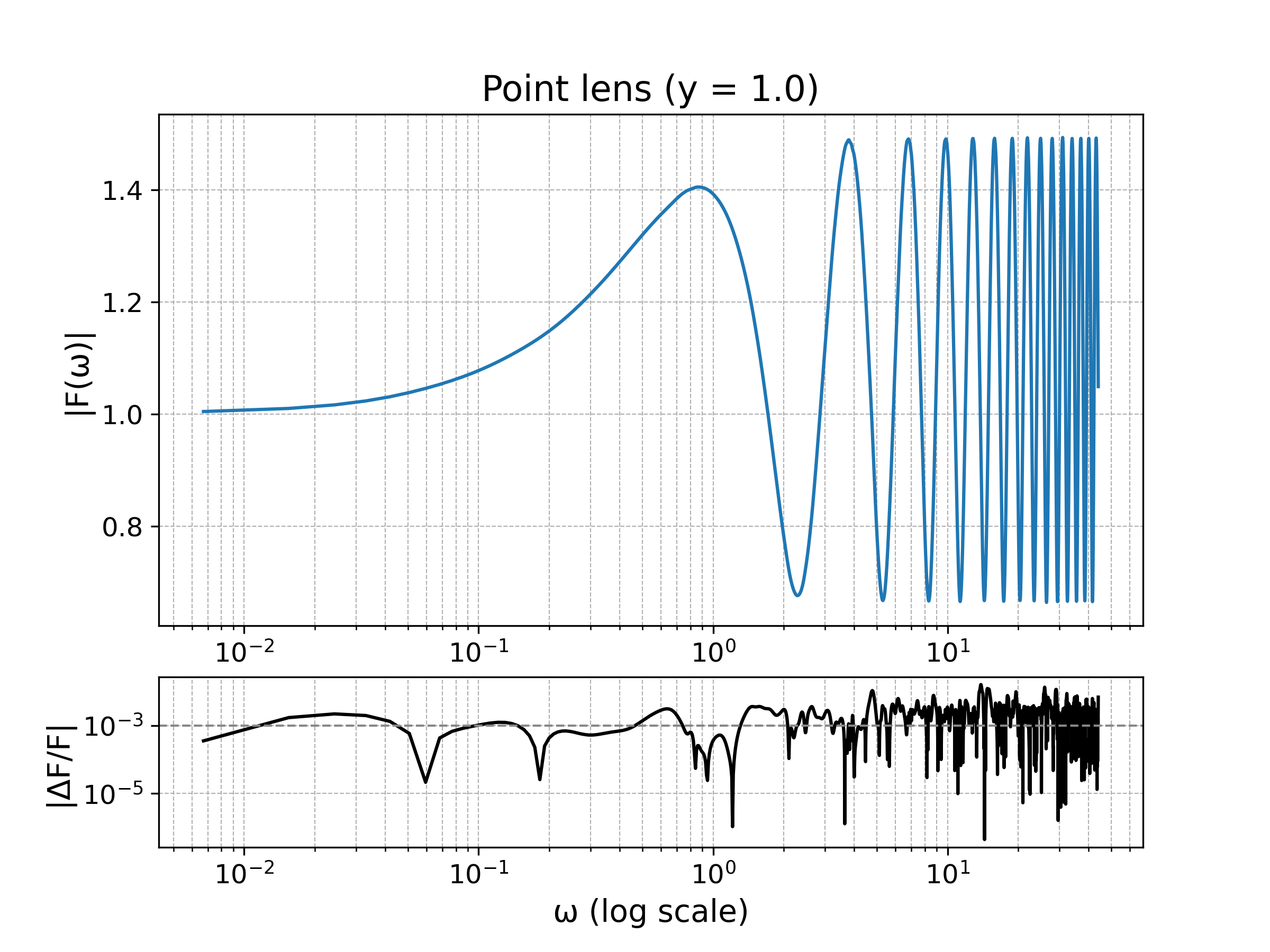} 
    \caption{$y=1.0$: Dense interference fringes with rapid phase evolution.}
    \label{fig:pred_y1.0}
  \end{subfigure}
  \caption{Model prediction vs. analytic truth. (\textit{a}) At $y=0.3$, the model captures the high-amplitude peak. (\textit{b}) At $y=1.0$, the model accurately resolves the transitions in oscillation frequency. The bottom panels show the per-frequency relative error, remaining consistently low ($\sim 10^{-3}$) in both cases.}
  \label{fig:predictions_examples}
\end{figure*}

\section{Results and Analysis} \label{sec:results}

\subsection{Overall Accuracy Assessment}

We quantify the model performance using the complex relative error evaluated on uniformly sampled random off-grid points $(\omega,y)$:
\[
\epsilon(\omega,y)=\frac{\bigl|F_{\rm pred}(\omega,y)-F_{\rm true}(\omega,y)\bigr|}{\bigl|F_{\rm true}(\omega,y)\bigr|}.
\]

As summarized in Table~\ref{tab:metrics}, our SIREN model attains per-mille level accuracy across the domain. The overall mean relative error is $1.10\times10^{-3}$, with a median of $8.30\times10^{-4}$. The 99th percentile error is \(4.94\times10^{-3}\), indicating well-controlled outliers.

The spatial distribution of errors is visualized in the heatmaps of Figure~\ref{fig:error_heatmaps}. We observe distinct error morphologies corresponding to different physical regimes.
At small impact parameters ($y \lesssim 1$), the relative errors are slightly elevated. This region corresponds to the quasi-geometric optics regime near the lens caustic, where the amplification factor exhibits a high dynamic range due to the constructive interference of two bright images. The sharp central peak ($|F| \gg 1$) introduces high-frequency gradients in the amplitude envelope, posing a challenge for the network to resolve simultaneously with the phase oscillations.
Conversely, in the large impact parameter regime ($y \gtrsim 3$), the system transitions into the weak-lensing diffraction tail. Here, the challenge shifts from amplitude dynamic range to phase frequency: the oscillation frequency scales as $\omega y$, leading to dense interference fringes. The domain decomposition strategy reduces the errors in this high-frequency limit. The SIREN model maintains a uniform accuracy profile across these physically distinct regimes, demonstrating robustness against both the amplitude singularities of strong lensing and the oscillatory catastrophes of weak lensing.

To provide a comprehensive view of the error behavior across the entire parameter space, particularly near the complex diffraction-interference transition, we visualize the relative error distribution in Fig.~\ref{fig:response_heatmap}. The parameter space is represented with a logarithmic axis for the dimensionless frequency $\omega$ and a linear axis for the impact parameter $y$. As illustrated, the model demonstrates remarkable robustness, with the majority of the parameter space generally maintaining relative errors at the $\mathcal{O}(10^{-3})$ level. Importantly, in the critical transition regime (moderate $y \sim 1$ and intermediate $\omega \sim 1\text{--}10$) where the amplification factor exhibits dense and rapid interference fringes, the SIREN architecture maintains high stability without introducing numerical artifacts. We observe a slight elevation in relative errors, occasionally approaching $\mathcal{O}(10^{-2})$, localized strictly in the very low-frequency regime ($\omega \in [10^{-2}, 10^{-1}]$). This minor localized deviation is primarily attributed to the relatively sparse sampling density within this specific sub-interval during our training phase. Refining the sampling strategy in this low-frequency regime will be a key direction for our future optimization work.

Figure~\ref{fig:response_heatmap} illustrates the statistical distribution of these errors. The histogram for off-grid points (Fig.~\ref{fig:hist_rand}) is heavily concentrated between $10^{-8}$ and $10^{-3}$. The Cumulative Distribution Function (CDF) confirms that over 99\% of test points achieve sub-percent accuracy. The training-grid statistics (Fig.~\ref{fig:hist_train}) closely mirror the off-grid results, indicating that the model performance on off-grid samples is consistent with the training error.

To verify the physical fidelity of the predictions, Figure~\ref{fig:predictions_examples} presents the waveforms at two representative impact parameters: $y=0.3$ (high dynamic range) and $y=1.0$ (intermediate regime). At $y=0.3$ (Fig.~\ref{fig:predictions_examples}a), the model accurately captures the pronounced central peak ($|F|\approx 2.5$) and the subsequent Airy pattern with relative errors consistently below $10^{-3}$. At $y=1.0$ (Fig.~\ref{fig:predictions_examples}b), the diffraction pattern transitions to a regime with a smoother envelope peak ($|F|\approx 1.4$) but increasingly rapid oscillations. The model tracks both the envelope and the phase of the oscillations robustly, maintaining per-mille level accuracy throughout the spectrum.

\subsection{Spectral Validation}
\label{subsec:spectral_validation}

\begin{figure*}[t]
  \centering
  \begin{subfigure}[t]{0.475\textwidth}
    \centering
    \includegraphics[width=\linewidth]{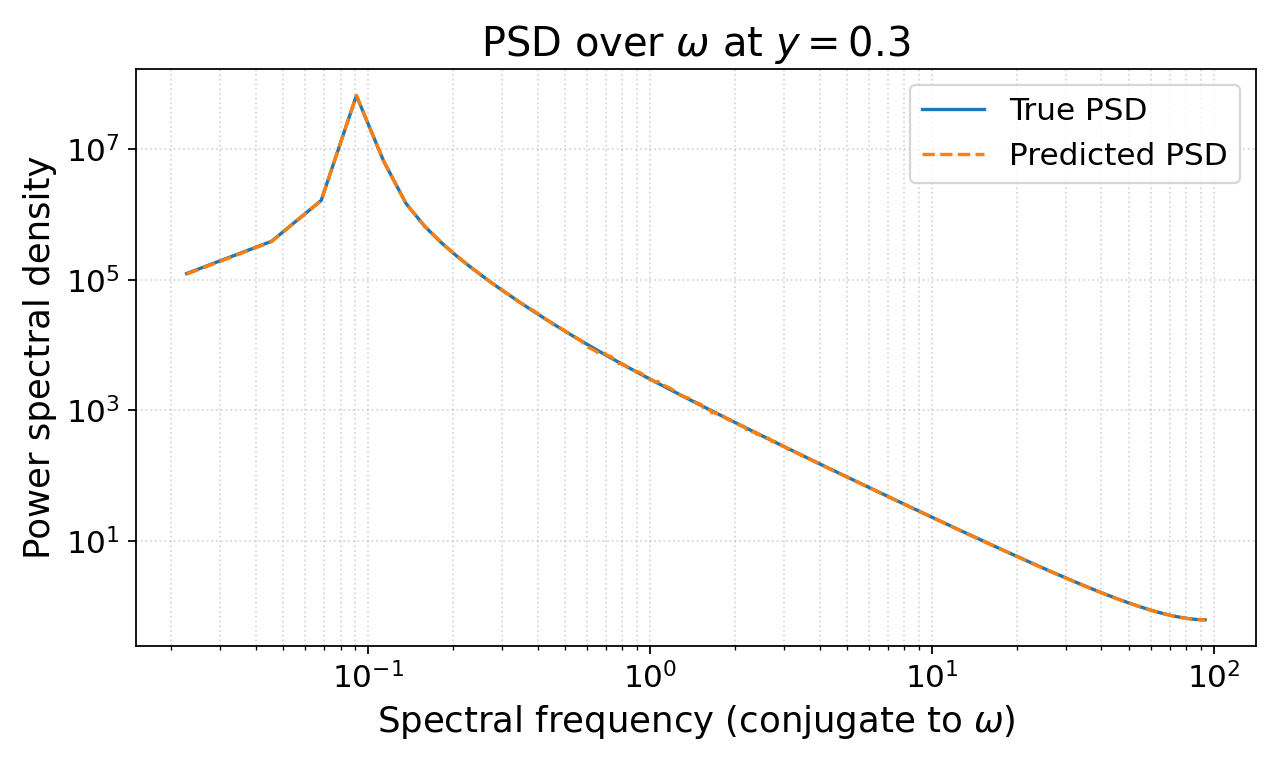}
    \caption{PSD at \(y=0.3\). The principal peak and high–\(\nu\) tail match closely.}
    \label{fig:psd_y0.3}
  \end{subfigure}\hfill
  \begin{subfigure}[t]{0.475\textwidth}
    \centering
    \includegraphics[width=\linewidth]{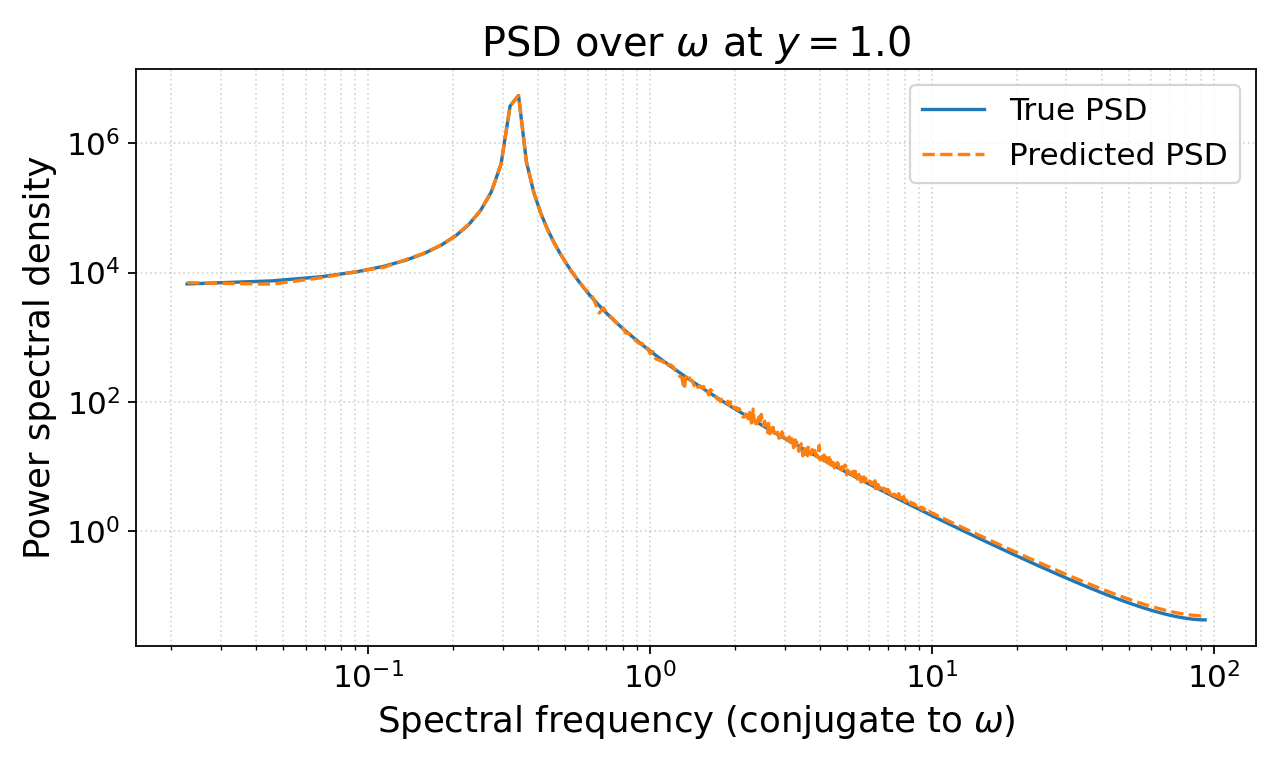}
    \caption{PSD at \(y=1.0\). Peak alignment and asymptotic slope are preserved.}
    \label{fig:psd_y1.0}
  \end{subfigure}
  \vspace{0.75ex}
  \caption{Comparison of Power Spectral Densities (PSD). The periodograms of the \emph{analytic} (solid) and \emph{model} (dashed) amplification factors \(F(\omega,y)\) are shown on log–log axes. The spectra are computed after de–meaning, Tukey windowing (\(\alpha=0.25\)), and zero–padding.}
  \label{fig:psd_comparison}
\end{figure*}

We assess the spectral fidelity of the model using representative impact parameters \(y=0.3\) and \(y=1.0\). For a fixed \(y\), we characterize the spectral content of the \emph{complex} amplification factor by its Power Spectral Density (PSD). We estimate the PSD using the periodogram, defined as the squared modulus of the discrete Fourier transform (DFT)~\citep{percival1993spectral}:
\[
P_F(\nu)=\bigl|\mathcal{F}_\omega\{F(\omega,y)\}\bigr|^{2},\qquad \nu\ge 0,
\]
where \(\mathcal{F}_\omega\) denotes the DFT applied to the discretized \(F(\omega,y)\) on a uniform \(\omega\)–grid, and \(\nu\) represents the spectral frequency conjugate to \(\omega\) (quantifying the rate of oscillation of the diffraction fringes).
Since we focus on the relative spectral structure, the standard normalization factor is omitted.
In practice, the series is mean-subtracted, multiplied by a Tukey window (\(\alpha=0.25\)) to mitigate spectral leakage, and zero–padded to the next power of two prior to the DFT. We verified that the qualitative conclusions are robust to moderate variations in the windowing and padding parameters.

Figure~\ref{fig:psd_comparison} compares the PSDs of the analytic PML solution and the model prediction.
At \(y=0.3\) (Fig.~\ref{fig:psd_y0.3}), the principal diffraction peak and the power–law tail at high \(\nu\) overlap over several orders of magnitude, indicating that the model preserves both the broadband envelope and the fine–scale oscillatory content.
At \(y=1.0\) (Fig.~\ref{fig:psd_y1.0}), the peak alignment and the asymptotic slope are accurately recovered.
While minor residuals appear at intermediate frequencies due to the dense fringe pattern, they do not deviate significantly from the ground truth scaling.
Overall, these tests confirm that the SIREN model successfully reproduces the multiscale spectral structure of \(F(\omega,y)\) across the relevant domain.

\subsection{Performance Comparison with NUMERICAL INTEGRATION METHODS}
\label{sec:zpae_comparison}

To assess the practical utility of our approach, we compare the SIREN model against the classical Zero Points Asymptotic Expansion (ZPAE) method (Sec.~\ref{sssec:zpae}), as well as the Asymptotic Expansion (AE) and Zero Points Integral (ZPI) methods, in terms of both accuracy and computational efficiency. Additionally, we include the high-precision analytical evaluation computed via the \texttt{mpmath} library to serve as our absolute ground-truth baseline.

\subsubsection{Accuracy Robustness}
\begin{figure}[h]
  \centering
  \includegraphics[width=\columnwidth]{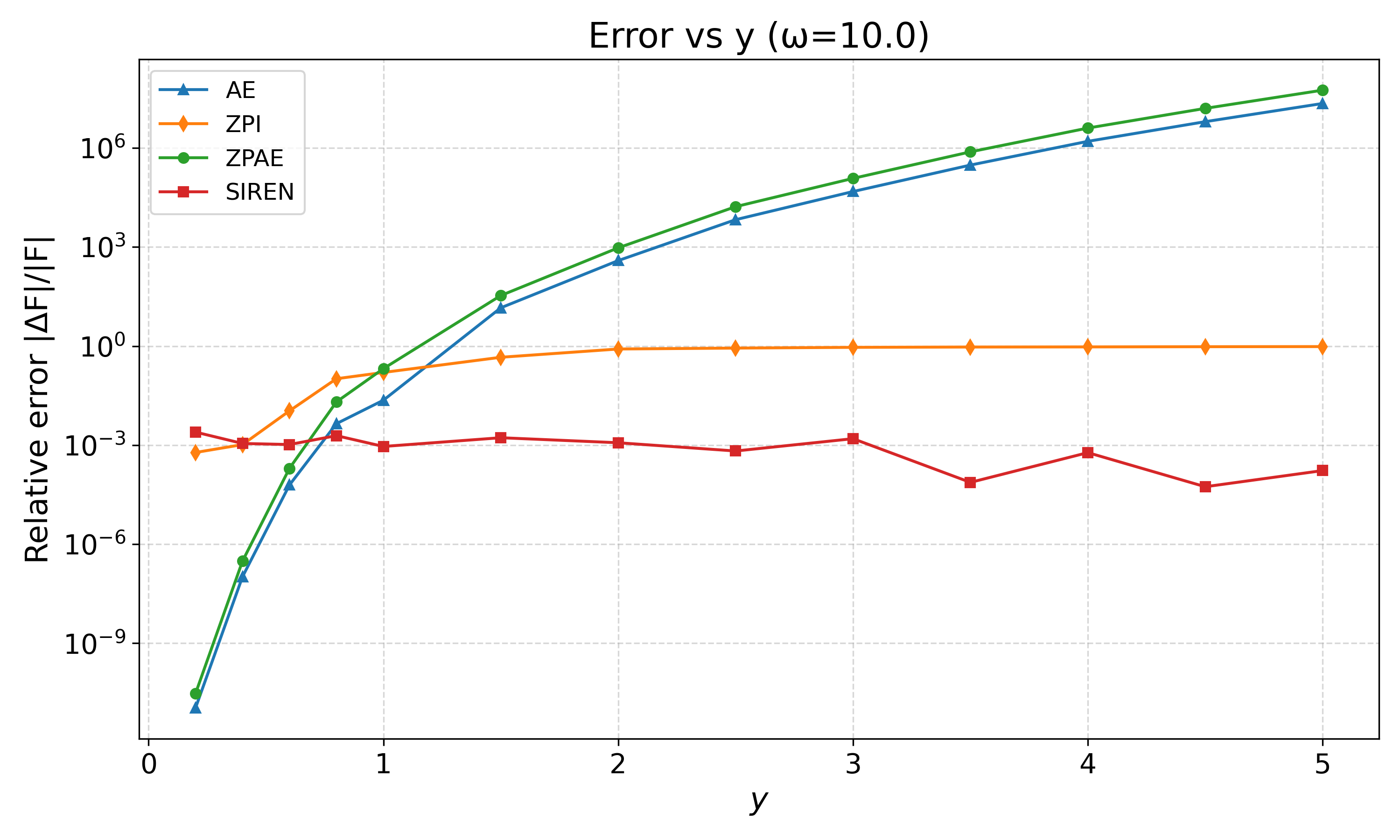}
  \caption{Relative error $\lvert\Delta F/ F\rvert$ versus $y$ at fixed $\omega=10.0$. The ZPAE and AE approximations achieve high accuracy only in the immediate vicinity of the expansion center ($y=0.2$) but deteriorate exponentially as $y$ deviates. The ZPI method is stable but yields higher baseline errors. In contrast, the SIREN model maintains stable $\sim10^{-3}$ accuracy across the entire domain.}
  \label{fig:zpae_vs_siren}
\end{figure}

We evaluate the accuracy using standard empirical parameters $k=5, \nu=7$ (and $m=15$ for ZPI) at a fixed frequency $\omega=10.0$. Figure~\ref{fig:zpae_vs_siren} compares the relative error $\lvert\Delta F\rvert/\lvert F\rvert$ as the impact parameter $y$ varies from 0.2 to 5.0. 

The ZPAE and AE methods exhibit extreme sensitivity to the alignment of their expansion centers. While they achieve vanishing residuals (below $10^{-8}$) specifically at the anchor point $y=0.2$, the approximation error diverges rapidly as $y$ deviates from this configuration. The error increases by orders of magnitude, so the methods require computationally expensive retuning to remain reliable. Meanwhile, the ZPI method, although more stable across the domain, yields a consistently high baseline error under standard truncation settings. In comparison, the SIREN model maintains a uniformly low relative error of order $10^{-3}$ across the entire domain. This confirms that the neural network successfully captures the global oscillatory behavior of the diffraction integral without requiring case-by-case hyperparameter adjustments.

\subsubsection{Computational Efficiency}
\label{sssec:efficiency}

To quantify inference latency, we benchmarked the SIREN model against the Mpmath, AE, ZPI, and ZPAE implementations by sampling across representative domains of both \(\omega\) and \(y\). 

\begin{figure*}[htbp]
  \centering
  \begin{subfigure}[t]{0.48\textwidth}
    \centering
    \includegraphics[width=\linewidth]{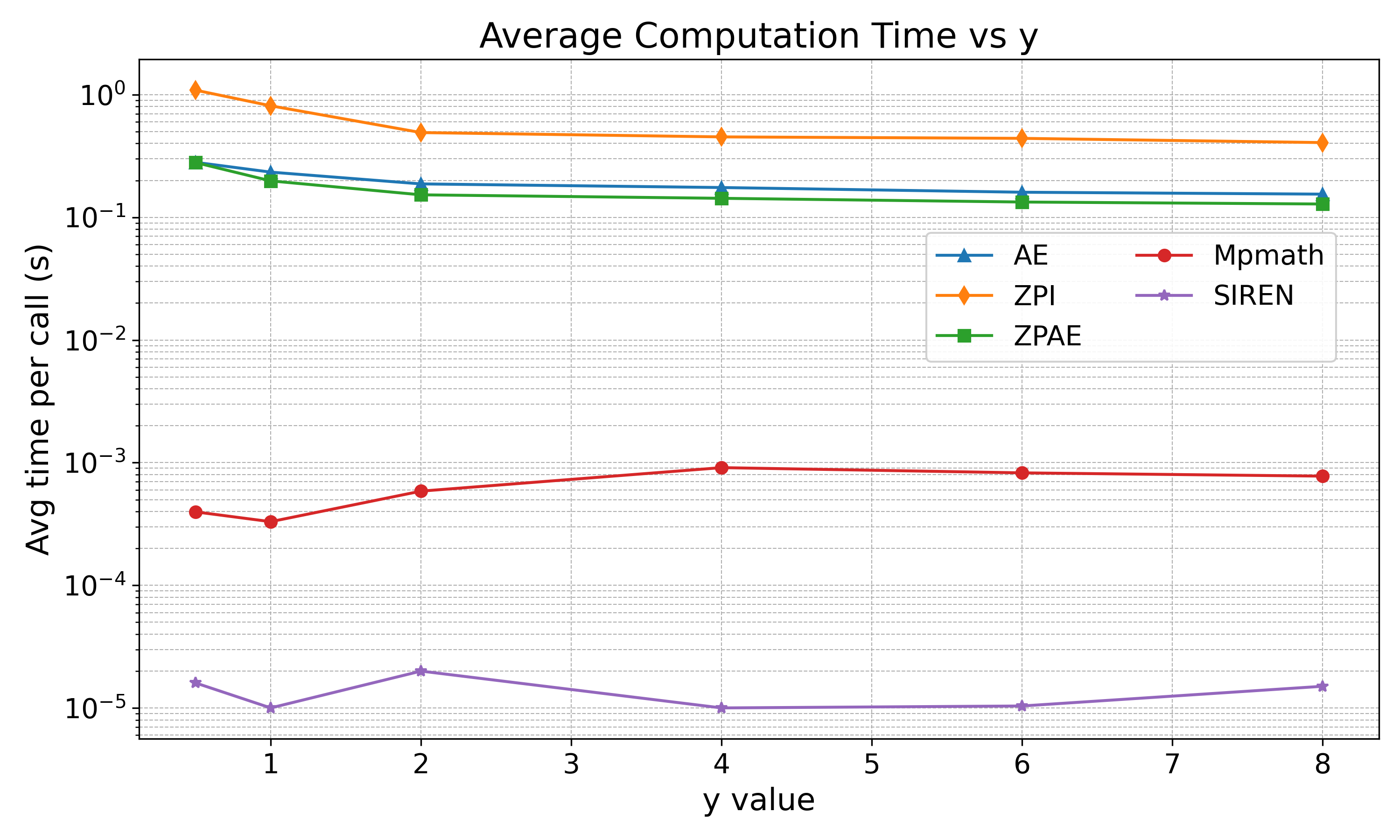}
    \caption{Average computation time vs $y$. The SIREN model is consistently faster by orders of magnitude.}
    \label{fig:time_vs_y}
  \end{subfigure}
  \hfill
  \begin{subfigure}[t]{0.48\textwidth}
    \centering
    \includegraphics[width=\linewidth]{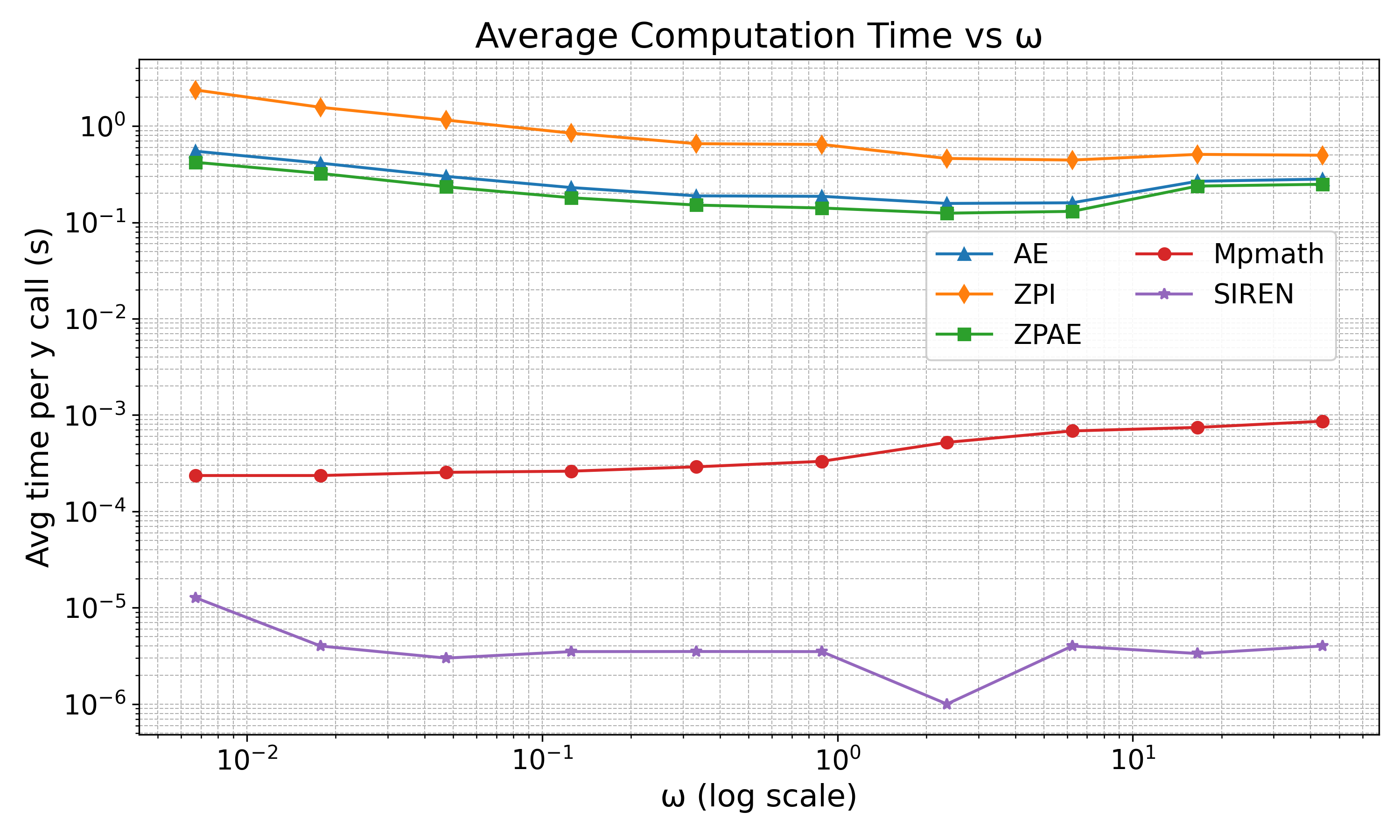}
    \caption{Average computation time vs $\omega$. SIREN maintains $\mathcal{O}(1)$ latency regardless of frequency.}
    \label{fig:time_vs_omega}
  \end{subfigure}
  \caption{Computational scaling comparison. The plots contrast the analytic integration (Mpmath), AE, ZPI, ZPAE, and the SIREN model. Note that the y-axis is in log scale. The SIREN model demonstrates the lowest latency ($\sim 10^{-5}$\,s) and exhibits constant-time complexity.}
  \label{fig:time_sweeps}
\end{figure*}

As illustrated in Fig.~\ref{fig:time_sweeps}, the SIREN model achieves a consistent latency of \(\sim 10^{-5}\)\,\text{s} per sample. This represents a substantial speedup, being approximately two orders of magnitude faster than the high-precision analytic Mpmath baseline (\(\sim 10^{-3}\)\,\text{s}), up to four orders faster than the unoptimized ZPAE and AE implementations (\(\sim 10^{-1}\)\,\text{s}), and five orders faster than the segmented ZPI method (\(\sim 10^{0}\)\,\text{s}).

Furthermore, the results in Fig.~\ref{fig:time_vs_y} and Fig.~\ref{fig:time_vs_omega} demonstrate the robustness of the neural estimator across the parameter space. The computational cost remains stable at this low latency level regardless of variations in the impact parameter \(y\) or the frequency \(\omega\). This consistent sub-millisecond efficiency is essential for supporting real-time lensing computations and large-scale parameter scans in GW data analysis.

\subsection{Phase Accuracy Analysis}

We decompose the amplification factor as
\[
F(\omega,y) = |F(\omega,y)|\,e^{i\phi(\omega,y)},
\]
and track its phase $\phi$. The phase structure is known to affect detectability in matched filtering and to introduce parameter degeneracies. Following the standard definition in circular statistics, we adopt the shortest arc distance as the phase error measure:
\[
\Delta\phi = \min\!\Bigl(\,|\phi_{\text{true}}-\phi_{\text{pred}}|\,,\, 
2\pi - |\phi_{\text{true}}-\phi_{\text{pred}}|\,\Bigr).
\]
Here, we additionally report the phase error to highlight the phase fidelity in the wave optics regime.

Table~\ref{tab:phase_accuracy} summarizes the mean, median, 95th-percentile, and maximum phase errors (in milliradians) for representative impact parameters in each of the four $y$-intervals.
Notably, the maximum phase error observed in our testing is approximately 15.36 mrad.
To understand this value, we compare it against the instrumental calibration uncertainty of current GW detectors. During the LIGO O3 observing run, the systematic phase calibration error was estimated to be approximately 2--5 degrees ($\sim 35$--$87$ mrad) in the sensitive frequency band.
Our model's residual phase error remains well below this instrumental floor.
Furthermore, according to waveform accuracy standards, a phase error of this magnitude corresponds to a negligible mismatch, ensuring that the neural network approximation does not introduce significant systematic bias into parameter estimation or reduce the recoverable signal-to-noise ratio.

\begin{table}[h]
\centering
\caption{Phase error statistics for representative \(y\) values in each domain interval (unit: mrad).}
\label{tab:phase_accuracy}
\small
\setlength{\tabcolsep}{4pt}
\begin{tabular}{lccccc}
\hline
Interval & \(y\)   & Mean   & Median & 95th \%ile & Maximum \\
\hline
\(\mathcal{I}_1\) &[0.2,1.0] & 1.43  & 1.10   & 3.93       & 15.36    \\
\(\mathcal{I}_2\) &[1.0,3.0]  & 1.33  & 1.05   & 3.46       & 9.15     \\
\(\mathcal{I}_3\) &[3.0,6.0]  & 0.26  & 0.22   & 0.61       & 7.56     \\
\(\mathcal{I}_4\) &[6.0,10.0] & 0.99  & 0.82   & 2.46       & 5.60     \\
\hline
\end{tabular}
\end{table}

\subsection{Generalization Capability: Extension to SIS Model}
\label{sec:sis_results}

To validate the generalization capability of our framework beyond the PML, we extended the pipeline to the SIS model. The primary objective was to verify that the SIREN architecture captures intrinsic wave-optics diffraction behavior rather than overfitting to a specific potential. We employed the identical network architecture and training strategy used for the PML baseline, modifying only the ground-truth generation module to incorporate the SIS diffraction integral~\citep{pagano2020lensinggw}. As shown in Fig.~\ref{fig:sis_predictions}, the model accurately reconstructs the spectral amplification for representative impact parameters (\(y=0.3\) and \(y=1.0\)), successfully capturing both the high-amplitude diffraction peaks and the dense interference fringes while maintaining relative errors generally below the \(10^{-3}\) level. To quantitatively verify this generalization, we evaluated the model on random test points strictly independent of the training grid. The error statistics and distributions are detailed in Appendix~\ref{sec:appendix_sis}. The SIS model achieves an overall mean relative error of \(2.29\times 10^{-4}\), maintaining stable accuracy across the entire evaluated parameter space. This robust performance on out-of-sample data further demonstrates the reliability of the architecture. This successful transferability implies that the proposed neural representation relies on the underlying wave physics rather than specific geometric symmetries, suggesting its broader applicability to more complex lens models.

\subsection{Generalization Capability: Extension to SIS Model}
\label{sec:sis_results}

\begin{figure*}[htbp]
  \centering
  \begin{subfigure}[t]{0.48\textwidth}
    \centering
    \includegraphics[width=\linewidth]{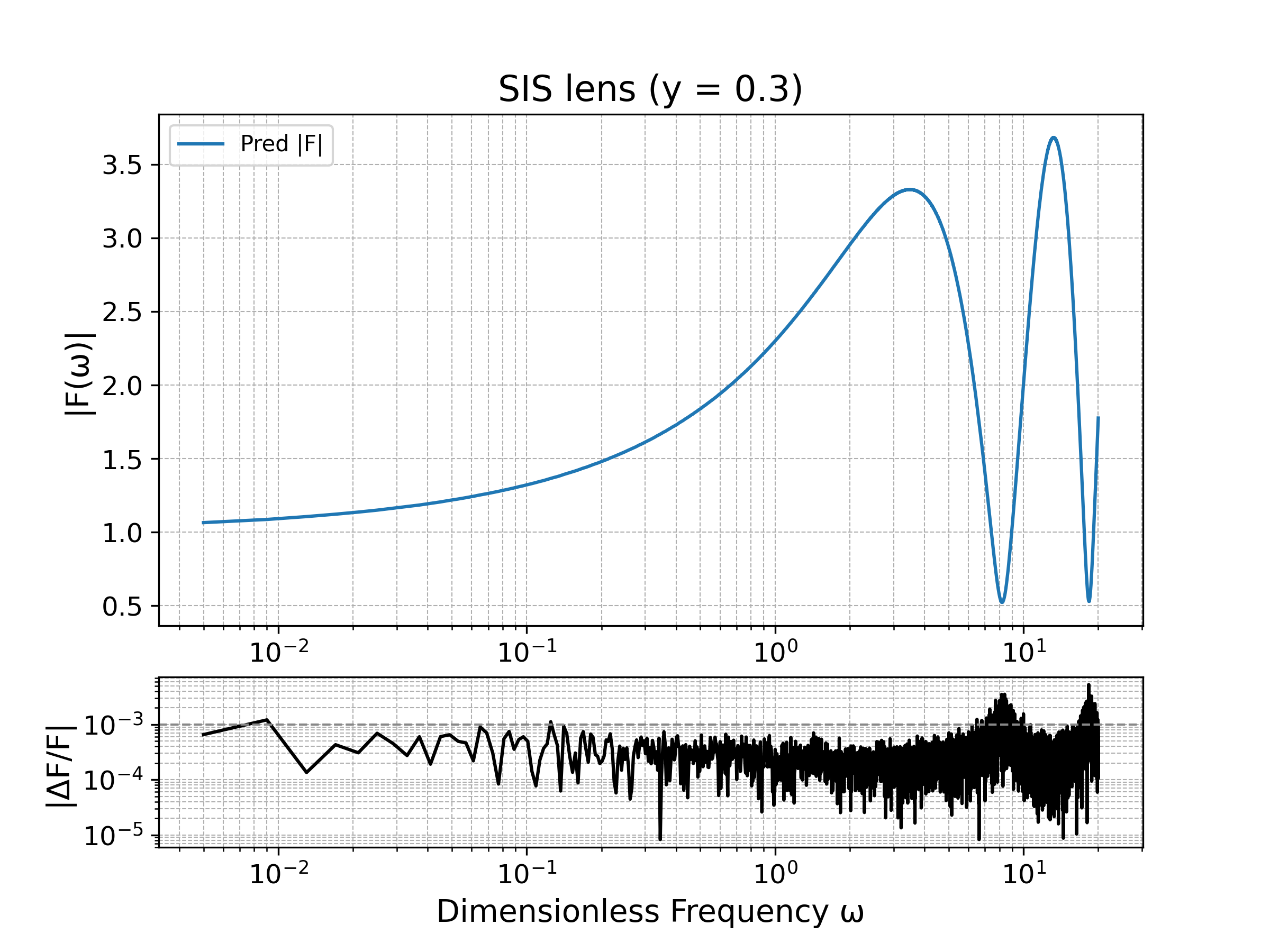}
    \caption{\(y=0.3\)}
    \label{fig:sis_y0.3}
  \end{subfigure}
  \hfill
  \begin{subfigure}[t]{0.48\textwidth}
    \centering
    \includegraphics[width=\linewidth]{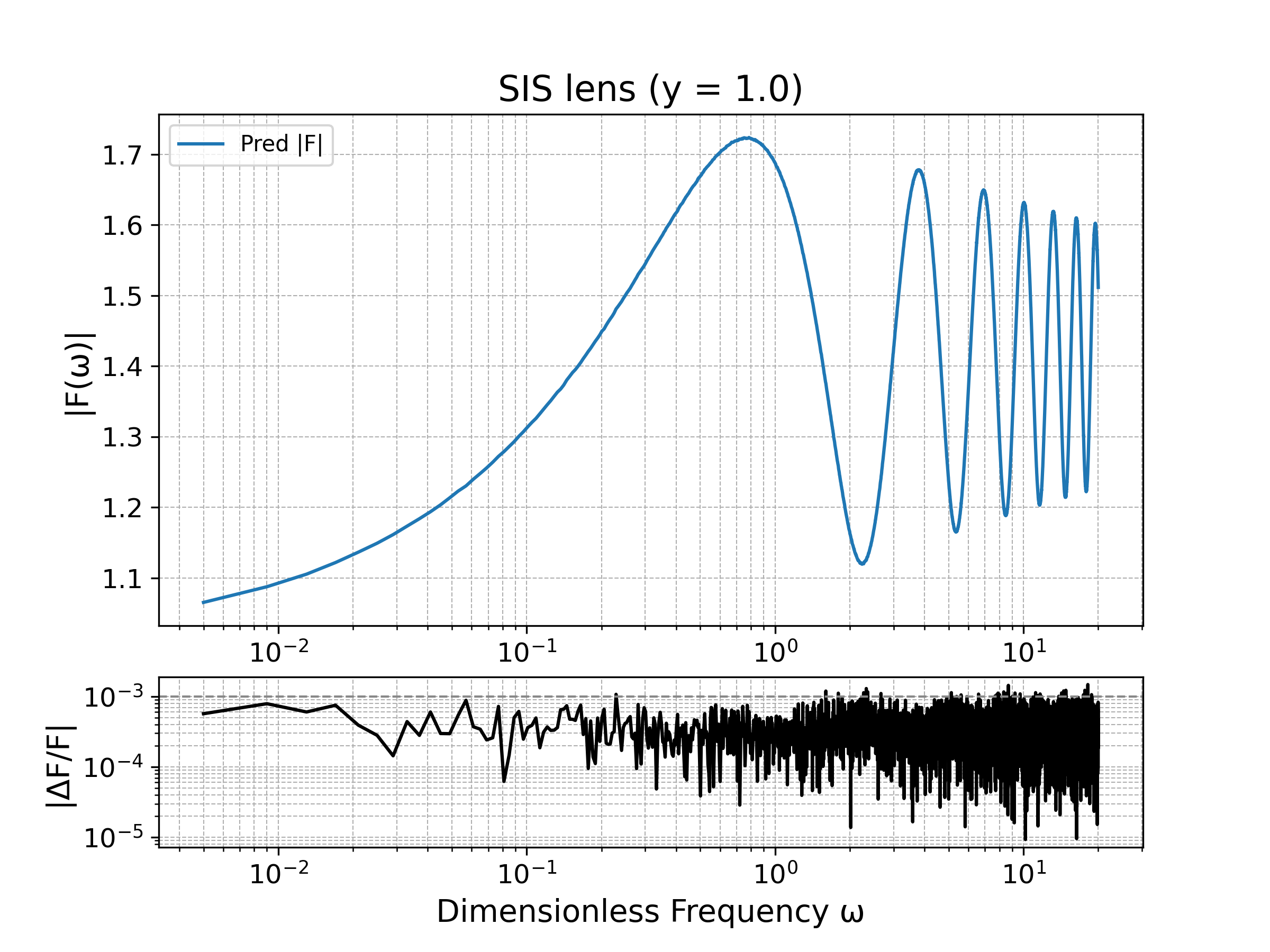} 
    \caption{\(y=1.0\)}
    \label{fig:sis_y1.0}
  \end{subfigure}
  \caption{Validation on the SIS model. The SIREN model accurately reconstructs the spectral amplification \(|F(\omega)|\) (top panels) and maintains low relative errors (bottom panels) for both strong lensing (\(y=0.3\)) and transition regimes (\(y=1.0\)), demonstrating the method's transferability.}
  \label{fig:sis_predictions}
\end{figure*}

To validate the generalization capability of our framework beyond the PML, we extended the pipeline to the SIS model. The primary objective was to verify that the SIREN architecture captures intrinsic wave-optics diffraction behavior rather than overfitting to a specific potential. We employed the identical network architecture and training strategy used for the PML baseline, modifying only the ground-truth generation module to incorporate the SIS diffraction integral~\citep{pagano2020lensinggw}. As shown in Fig.~\ref{fig:sis_predictions}, the model accurately reconstructs the spectral amplification for representative impact parameters (\(y=0.3\) and \(y=1.0\)). To quantitatively verify this generalization, we evaluated the model on random test points strictly independent of the training grid. The error statistics and distributions are detailed in Appendix~\ref{sec:appendix_sis}. The SIS model achieves an overall mean relative error of \(2.29\times 10^{-4}\), maintaining stable accuracy across the entire evaluated parameter space. This robust performance on out-of-sample data further demonstrates the reliability of the architecture. This successful transferability implies that the proposed neural representation relies on the underlying wave physics rather than specific geometric symmetries, suggesting its broader applicability to more complex lens models.

\section{Conclusions}
\label{sec:conclusions}

This study establishes a neural representation framework based on SIRENs to resolve computational bottlenecks in wave-optics lensing.
The network's periodic activation functions naturally match the diffraction integral's oscillatory kernel, enabling the model to capture high-frequency spectral features.
Quantitatively, the method achieves a relative accuracy of $\mathcal{O}(10^{-3})$ and speeds up computation by $\sim 100\times$ compared to numerical integration.
Concurrently, the framework shifts the heavy computational burden from ``online integration'' to ``offline training,'' resulting in $\mathcal{O}(1)$ inference complexity that remains constant regardless of the oscillation frequency. This constant-time inference performance offers a potential solution to the computational bottleneck in Bayesian parameter estimation pipelines. By providing fast and accurate likelihood evaluations with a stable execution time that remains independent of parameter variations, our framework supports a practical pathway toward tractable wave-optics parameter estimation for future gravitational wave data analysis.

A key feature of the model is its dimensionless formulation, which ensures intrinsic scale invariance.
While we validated the framework using stellar-mass lenses typical of LVK observations, this design allows direct application to supermassive black hole lensing in the LISA band without retraining.
High efficiency is essential for third-generation detectors, such as the Einstein Telescope (ET) \citep{2020JCAP...03..050M} and Cosmic Explorer (CE) \citep{2019BAAS...51g..35R}.
These instruments will likely detect hundreds of high-SNR lensing events annually \citep{2015JCAP...12..006D, 2022MNRAS.509.3772Y, 2023MNRAS.526..682G}.
Such a data volume is too high for traditional online integration methods, making our approach a necessary alternative.

In summary, coordinate-based neural representations provide a robust tool for precision astronomy.
Beyond gravitational lensing, matching neural architectures to physical kernels offers an effective path to accelerate the analysis of highly oscillatory systems in broader physical contexts.

\section*{Acknowledgments}
%We would like to thank the referees for their valuable comments, which considerably improved the original text.  
FZ would like to thank the MIT LIGO Laboratory for its continuous support and advice. FZ is supported by the National Natural Science Foundation of China (Grant No. 62372409) and the Ministry of Science and Technology of the People’s Republic of China (Grant No. 2023ZD0120704 of Project No. 2023ZD0120700).

\appendix

\section{Per-Interval Error Distributions} 

% —— Interval I1 ——  
\begin{figure*}[htbp]
  \centering
  \begin{subfigure}[t]{\textwidth}
    \centering
    \includegraphics[height=0.155\textheight]{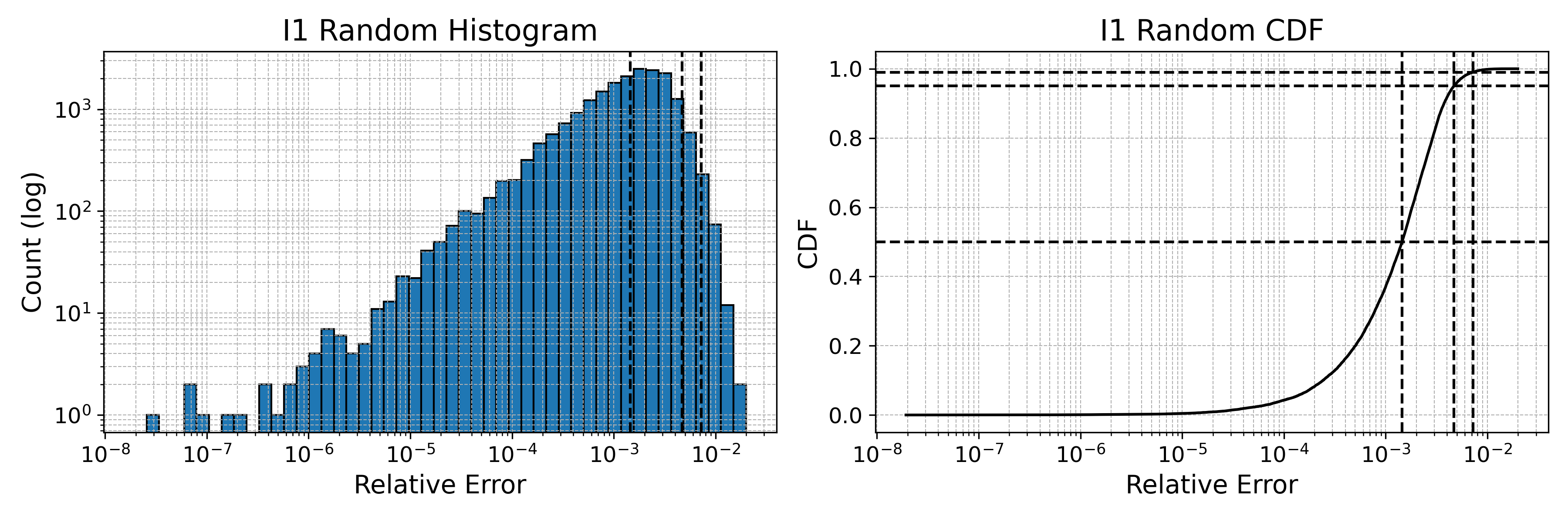}
    \caption{Interval \(\mathcal I_1\) (\(y\in[0.2,1.0]\)), random test points: relative‐error histogram (left) and CDF (right).}
    \label{fig:I1_rand}
  \end{subfigure}\\[1ex]
  \begin{subfigure}[t]{\textwidth}
    \centering
    \includegraphics[height=0.155\textheight]{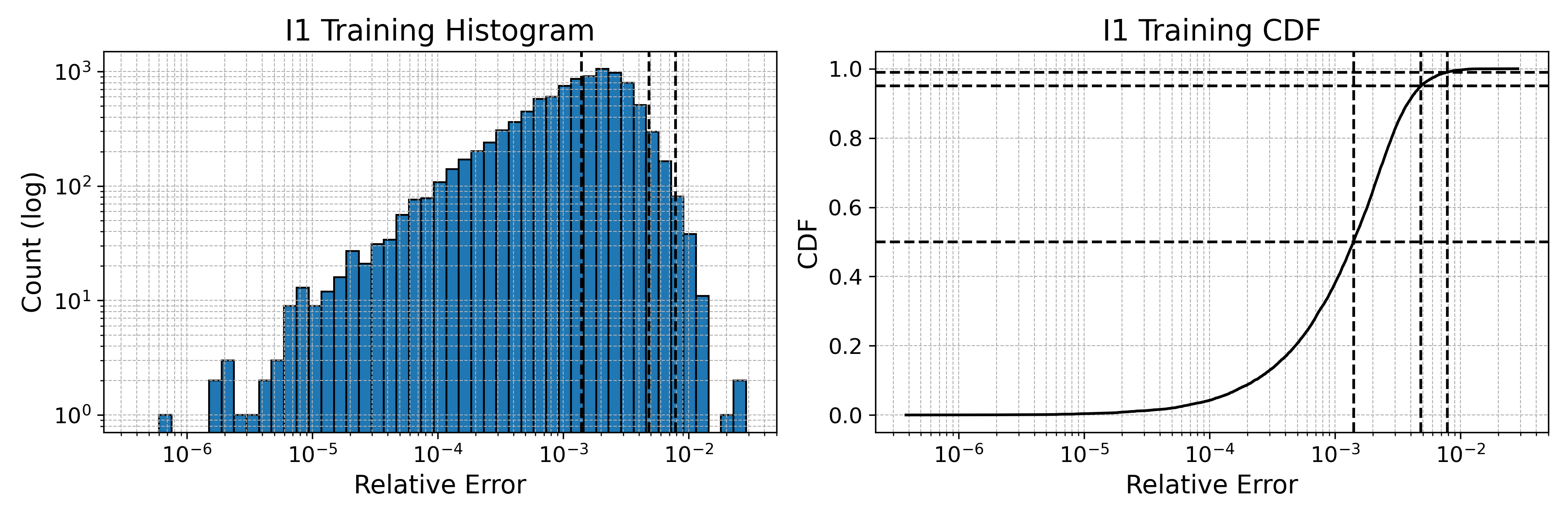}
    \caption{Interval \(\mathcal I_1\) (\(y\in[0.2,1.0]\)), training‐grid samples: relative‐error histogram (left) and CDF (right).}
    \label{fig:I1_train}
  \end{subfigure}
  \caption{Error distributions in the smallest‐\(y\) interval. Panel (a) shows the distribution over 20,000 random \((\omega,y)\) samples, while panel (b) shows the same statistics on the original training grid. Histograms (left halves) use log‐scaled counts and bins; CDFs (right halves) mark the 1\%, 5\%, 10\%, and 90\% quantiles.}
  \label{fig:app_I1}
\end{figure*}

% —— Interval I2 ——  
\begin{figure*}[htbp]
  \centering
  \begin{subfigure}[t]{\textwidth}
    \centering
    \includegraphics[height=0.155\textheight]{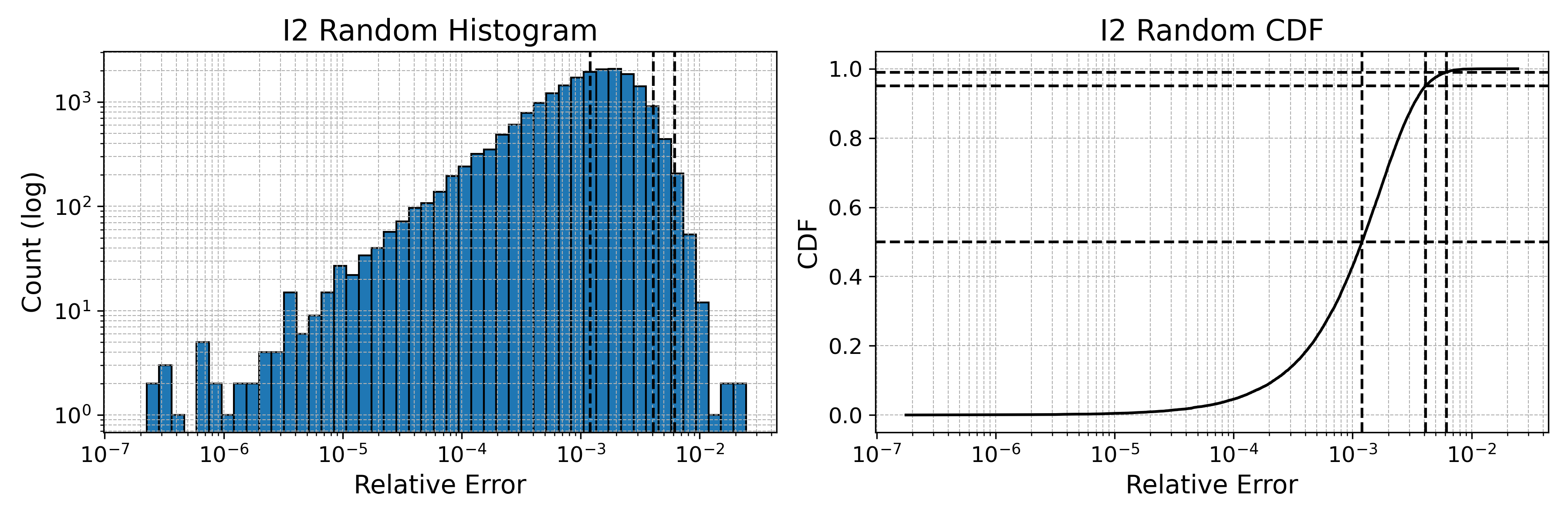}
    \caption{Interval \(\mathcal I_2\) (\(y\in[1.0,3.0]\)), random test points: histogram and CDF.}
    \label{fig:I2_rand}
  \end{subfigure}\\[1ex]
  \begin{subfigure}[t]{\textwidth}
    \centering
    \includegraphics[height=0.155\textheight]{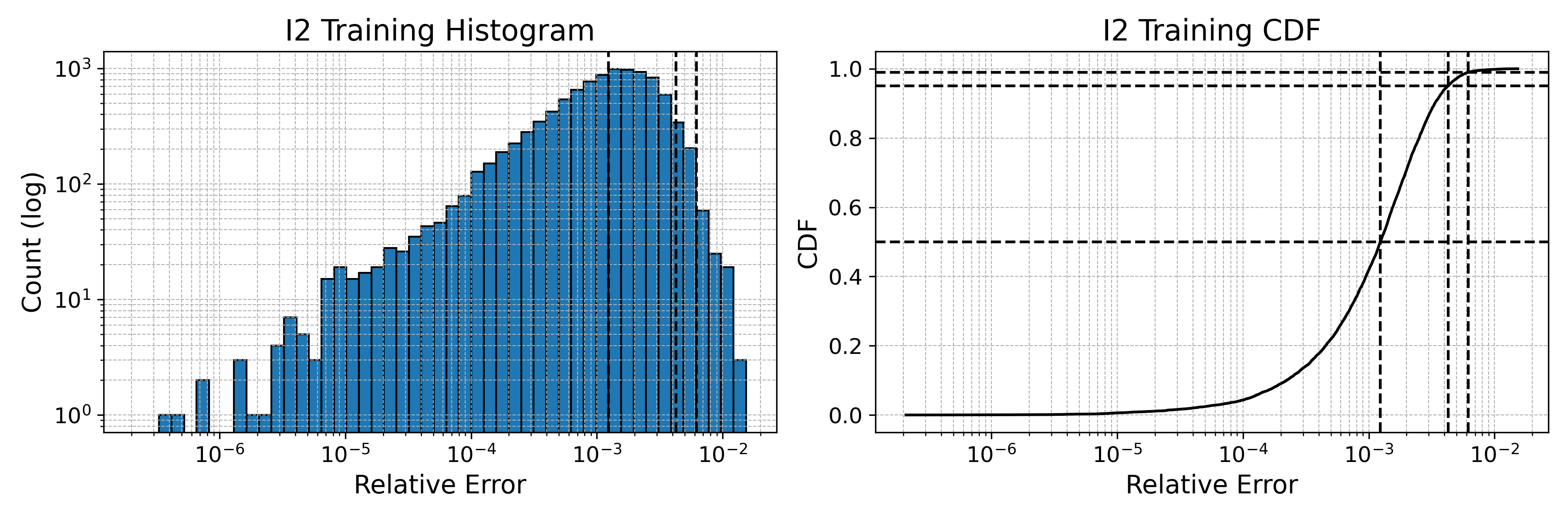}
    \caption{Interval \(\mathcal I_2\) (\(y\in[1.0,3.0]\)), training‐grid samples: histogram and CDF.}
    \label{fig:I2_train}
  \end{subfigure}
  \caption{Error distributions in the low‐to‐moderate‐\(y\) interval. Random‐point (a) and grid‐point (b) errors both concentrate in the \(10^{-7}\)–\(10^{-3}\) range, with over 90\% of samples below \(10^{-4}\) error.}
  \label{fig:app_I2}
\end{figure*}

% —— Interval I3 ——  
\begin{figure*}[htbp]
  \centering
  \begin{subfigure}[t]{\textwidth}
    \centering
    \includegraphics[height=0.155\textheight]{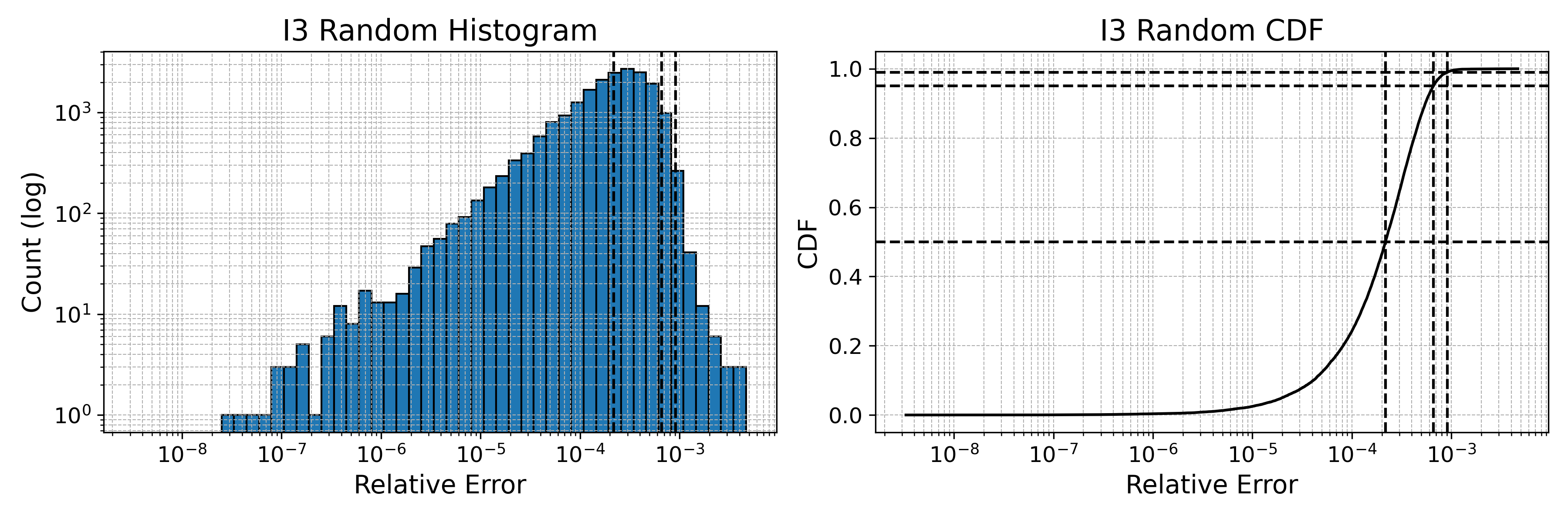}
    \caption{Interval \(\mathcal I_3\) (\(y\in[3.0,6.0]\)), random test points: histogram and CDF.}
    \label{fig:I3_rand}
  \end{subfigure}\\[1ex]
  \begin{subfigure}[t]{\textwidth}
    \centering
    \includegraphics[height=0.155\textheight]{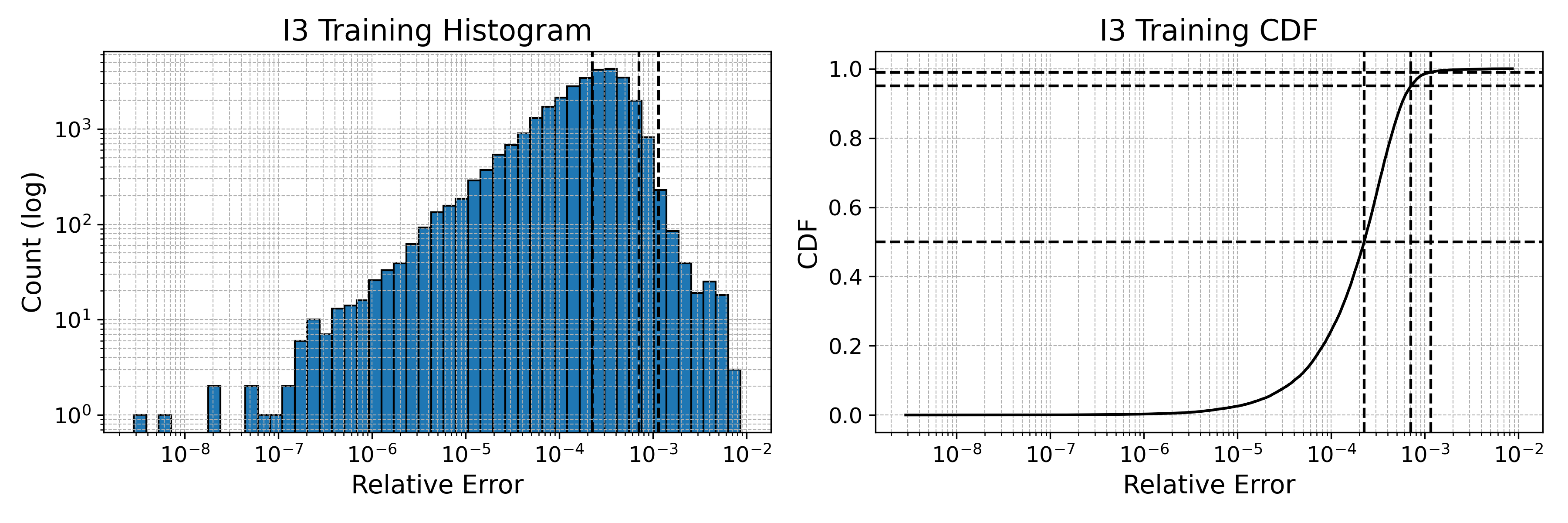}
    \caption{Interval \(\mathcal I_3\) (\(y\in[3.0,6.0]\)), training‐grid samples: histogram and CDF.}
    \label{fig:I3_train}
  \end{subfigure}
  \caption{Error distributions in the moderate‐\(y\) interval. Both random and grid CDFs (panels a and b) exceed 97\% below \(10^{-5}\) error, indicating peak network accuracy in this middle interval.}
  \label{fig:app_I3}
\end{figure*}

% —— Interval I4 ——  
\begin{figure*}[htbp]
  \centering
  \begin{subfigure}[t]{\textwidth}
    \centering
    \includegraphics[height=0.155\textheight]{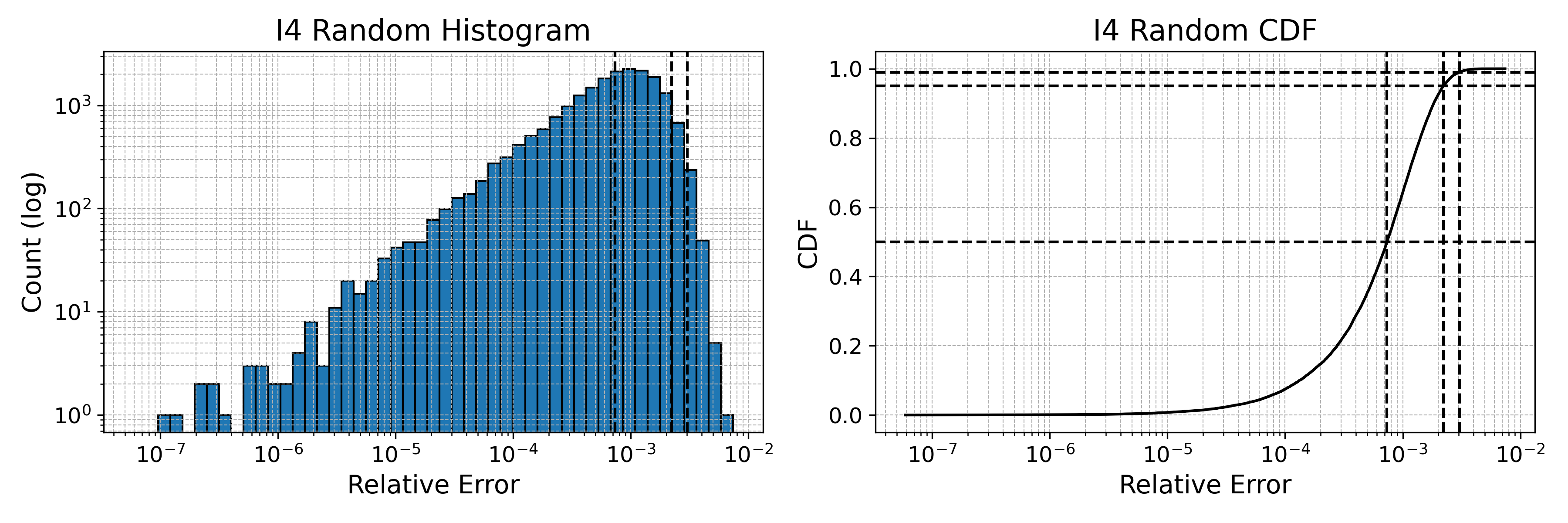}
    \caption{Interval \(\mathcal I_4\) (\(y\in[6.0,10.0]\)), random test points: histogram and CDF.}
    \label{fig:I4_rand}
  \end{subfigure}\\[1ex]
  \begin{subfigure}[t]{\textwidth}
    \centering
    \includegraphics[height=0.155\textheight]{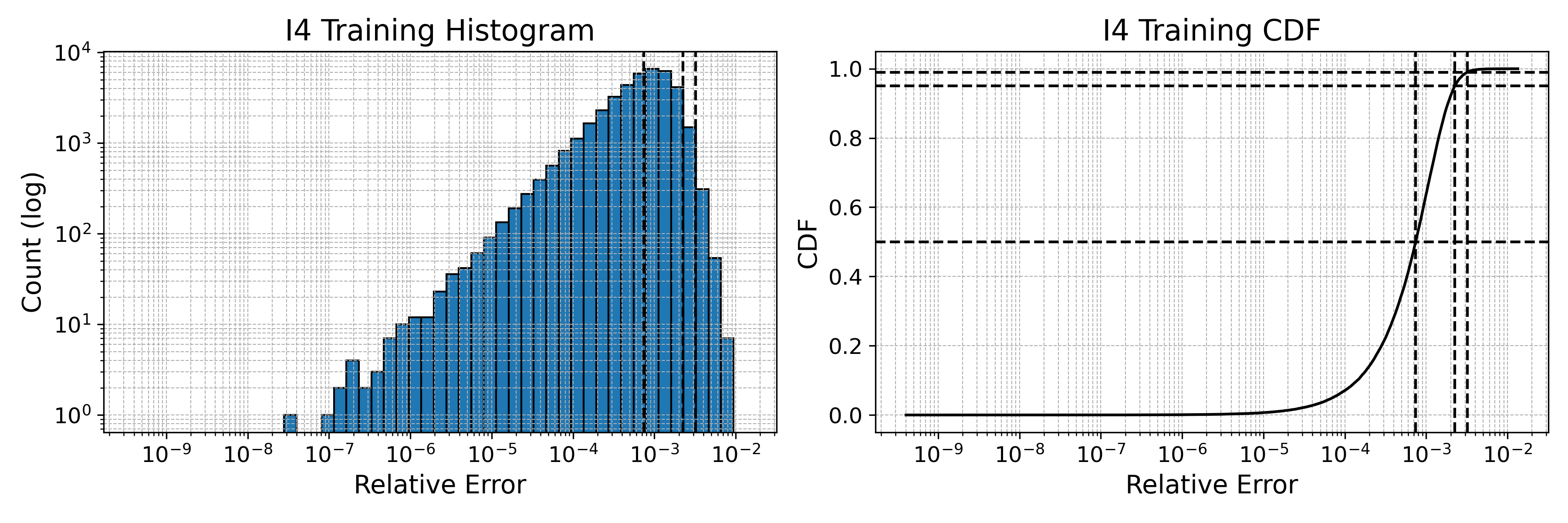}
    \caption{Interval \(\mathcal I_4\) (\(y\in[6.0,10.0]\)), training‐grid samples: histogram and CDF.}
    \label{fig:I4_train}
  \end{subfigure}
  \caption{Error distributions in the largest‐\(y\) interval. Even in this high‐frequency interval, over 95\% of points in both random (a) and grid (b) cases remain below \(10^{-4}\) error.}
  \label{fig:app_I4}
\end{figure*}

This appendix provides detailed histograms and CDFs for each of the four impact parameter intervals discussed in the main text.

The per–interval histograms and CDFs in Figs.~\ref{fig:I1_rand} and \ref{fig:I1_train} (I1, \(y\in[0.2,1.0]\)) show that off–grid errors are concentrated toward the higher end of our overall range, with a visible upper tail. The training–grid distribution closely matches the off–grid case and shows a slightly lower center, indicating comparable accuracy and stable generalization in the near–alignment band.

Figs.~\ref{fig:I2_rand} and \ref{fig:I2_train} (I2, \(y\in[1.0,3.0]\)) display a clear leftward shift relative to I1. Most probability mass moves toward smaller errors while the CDF rises more steeply, and the training–grid curves remain well aligned with the off–grid curves. This behavior reflects improved accuracy as \(y\) moves away from near alignment.

Figs.~\ref{fig:I3_rand} and \ref{fig:I3_train} (I3, \(y\in[3.0,6.0]\)) present the tightest distributions among all intervals. The histogram peak lies at the low–error end and the CDF shows a rapid rise, with very weak upper tails. Off–grid and training–grid results are nearly indistinguishable, indicating the highest accuracy and the strongest tail suppression in this intermediate band.

Figs.~\ref{fig:I4_rand} and \ref{fig:I4_train} (I4, \(y\in[6.0,10.0]\)) remain close to I3 in both scale and shape. Off–grid and training–grid distributions again agree well. A mild elevation is visible only in a small portion of the domain associated with large \(y\) and high \(\omega\), while the bulk of the mass remains at per–mille–level errors.

Across all intervals the off–grid statistics mirror the training–grid statistics, and the ordering I1 \(>\) I2 \(>\) I3 \(\approx\) I4 is consistent with the summary in Table~\ref{tab:metrics}. Errors are relatively higher for small \(y\), reach their lowest levels for \(3\le y\le6\), and increase slightly only near the high–\(y\), high–\(\omega\) corner. Overall, the model maintains per–mille–level accuracy with fast upper–tail decay and consistent generalization beyond the training points. For quantitative summaries, including mean, median, and upper–tail percentiles, see Table~\ref{tab:metrics}.

\section{SIS Model Error Distributions}
\label{sec:appendix_sis}

We provide detailed quantitative metrics for the SIS model evaluation to demonstrate the robustness of our architecture across different physical regimes. The model was tested on random off-grid points across four intervals of the impact parameter \(y\). Table~\ref{tab:sis_metrics} summarizes the complex relative error performance. The overall mean relative error is \(2.29\times 10^{-4}\), confirming the high accuracy of the SIREN architecture on the SIS potential. 

Consistent with the PML baseline, the performance breakdown reveals that the most challenging region is the innermost interval \(\mathcal{I}_{1} [0.2, 1.0)\). Here, the amplification factor exhibits sharp amplitude gradients and high dynamic range due to strong lensing and caustic proximity, resulting in a slightly elevated mean error of \(1.25\times 10^{-3}\). Conversely, for moderate to large impact parameters (\(\mathcal{I}_{2}\) to \(\mathcal{I}_{4}\)) where dense wave-optics interference fringes dominate, the SIREN architecture perfectly resolves the rapid phase oscillations, suppressing the mean relative errors to the \(10^{-5}\) and \(10^{-4}\) levels.

\begin{table}[htbp]
\centering
\caption{Summary of the SIS model relative error performance. Metrics include mean, median, 95th/99th percentiles, and maximum error evaluated on random test samples.}
\label{tab:sis_metrics}
\begin{tabular}{lccccc}
\hline
Metric & Overall & \(\mathcal{I}_{1} [0.2, 1.0)\) & \(\mathcal{I}_{2} [1.0, 3.0)\) & \(\mathcal{I}_{3} [3.0, 6.0)\) & \(\mathcal{I}_{4} [6.0, 10.0]\) \\
\hline
Mean Rel. Error   & \(2.29\times 10^{-4}\) & \(1.25\times 10^{-3}\) & \(1.00\times 10^{-4}\) & \(6.51\times 10^{-5}\) & \(1.60\times 10^{-4}\) \\
Median Rel. Error & \(8.83\times 10^{-5}\) & \(3.83\times 10^{-4}\) & \(6.94\times 10^{-5}\) & \(4.97\times 10^{-5}\) & \(1.26\times 10^{-4}\) \\
95th Percentile   & \(4.91\times 10^{-4}\) & \(3.58\times 10^{-3}\) & \(2.22\times 10^{-4}\) & \(1.83\times 10^{-4}\) & \(4.10\times 10^{-4}\) \\
99th Percentile   & \(1.30\times 10^{-3}\) & \(8.08\times 10^{-3}\) & \(7.24\times 10^{-4}\) & \(2.46\times 10^{-4}\) & \(6.01\times 10^{-4}\) \\
Maximum Error     & \(1.97\times 10^{-2}\) & \(1.97\times 10^{-2}\) & \(7.20\times 10^{-3}\) & \(5.91\times 10^{-3}\) & \(4.07\times 10^{-3}\) \\
\hline
\end{tabular}
\end{table}

\begin{figure}[htbp]
  \centering
  \includegraphics[width=0.8\linewidth]{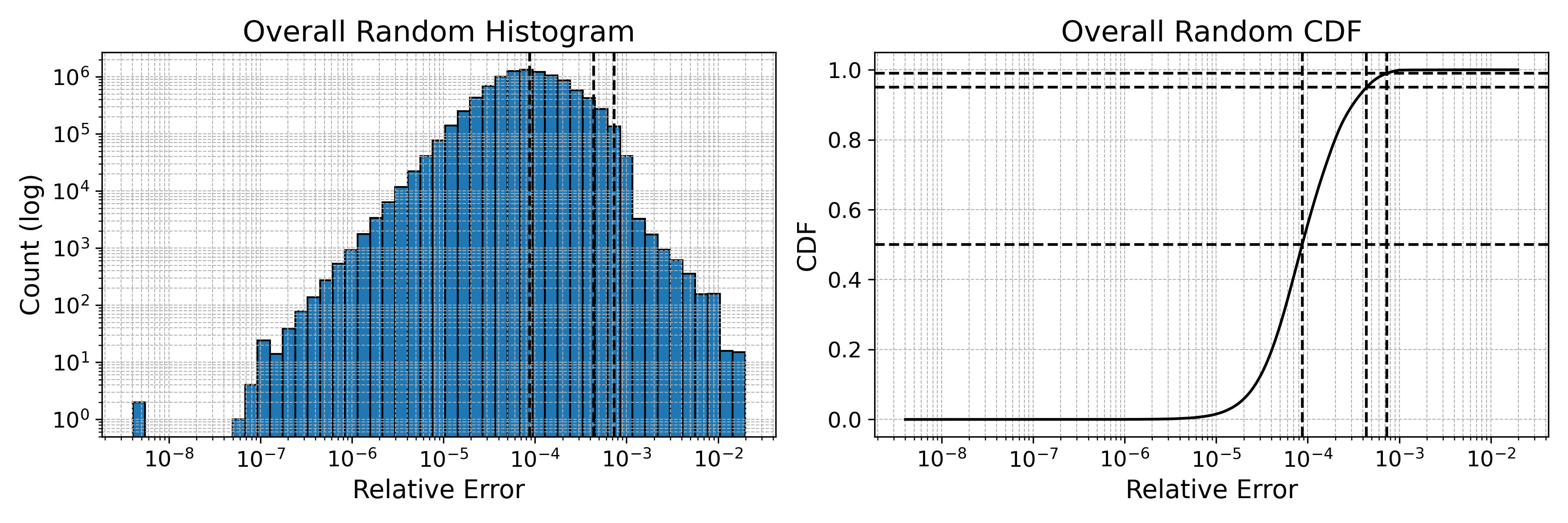}
  \caption{Overall distributions of relative error for the SIS model on random test points. The histogram (left) and CDF (right) demonstrate stable generalization and bounded extreme errors.}
  \label{fig:sis_histogram}
\end{figure}

Figure~\ref{fig:sis_histogram} further visualizes this stability by illustrating the statistical distribution of the overall relative errors. The histogram (left panel) is heavily concentrated around \(10^{-4}\) with a rapid decay in the upper tail. The Cumulative Distribution Function (CDF, right panel) explicitly confirms that 95\% of the unseen test points achieve an error below \(4.91\times 10^{-4}\), and 99\% are strictly bounded below \(1.30\times 10^{-3}\). The absence of catastrophic outliers in this distribution matches the strong generalization behavior observed in the PML baseline, verifying that the neural framework reliably captures the underlying diffraction integral rather than overfitting.

\clearpage
\twocolumngrid
% \bibliography{reference}% Produces the bibliography via BibTeX.
\bibliography{reference}{}
\bibliographystyle{aasjournalv7}

\end{document}